\documentclass[twocolumn]{aastex631}

\usepackage{amsmath}
\usepackage{xcolor}

\usepackage[T1]{fontenc}

\usepackage{ulem}

\usepackage{esint}

\newcommand{\be}{\begin{equation}}
\newcommand{\ee}{\end{equation}}
\newcommand{\bea}{\setlength\arraycolsep{2pt} \begin{eqnarray}}
\newcommand{\eea}{\end{eqnarray}}
\newcommand{\nn}{\nonumber}

\newcommand{\Mc}[1]{\mathcal{#1}}

\newcommand{\mO}{{\mathcal O}}

\newcommand{\mP}{{\mathcal P}}

\newcommand{\mE}{{\mathcal E}}
\newcommand{\mL}{{\mathcal L}}

\newcommand{\md}{\mathrm{d}}

\def\a{\alpha}
\def\b{\beta}
\def\c{\chi}
\def\d{\delta}
\def\D{\Delta}
\def\f{\frac}
\def\g{\gamma}

\def\G{\Gamma}

\def\l{\left}
\def\L{\Lambda}
\def\m{\mu} 
\def\n{\nu} 
\def\nn{\nonumber}

\def\p{\phi} 
 
\def\r{\right}
\def\s{\sigma}
\def\S{\Sigma}
\def\t{\theta}
\def\T{\Theta}
\def\k{\kappa}

\def\o{\omega}

\def\be{\begin{equation}}
\def\ee{\end{equation}}
\def\bag{\begin{aligned}}
\def\eag{\end{aligned}}
\def\bea{\begin{eqnarray}}
\def\eea{\end{eqnarray}}
\def\ba{\begin{array}}
\def\ea{\end{array}}

\def\bc{\begin{center}}
\def\ec{\end{center}}

\begin{document}

\title{Polarization Architecture of Steady GRMHD Jets from the Horizon to Infinity}

\correspondingauthor{Yehui Hou}
\email{yehuihou@sjtu.edu.cn}
\correspondingauthor{Bin Chen}
\email{chenbin1@nbu.edu.cn}

\author[0000-0003-0869-4601]{Zhenyu Zhang}
\affiliation{Institute of Fundamental Physics and Quantum Technology, \& School of Physical Science and Technology, \\Ningbo University, Ningbo, Zhejiang 315211, P. R. China}

\author[0000-0002-9434-3930]{Yehui Hou}
\affiliation{Tsung-Dao Lee Institute, Shanghai Jiao-Tong University, Shanghai, 201210, P. R. China}

\author[0009-0007-0936-3747]{Yu Song}
\affiliation{School of Physics, Peking University, No.5 Yiheyuan Rd, Beijing 100871, P. R. China}
\affiliation{Shanghai Astronomical Observatory, Chinese Academy of Sciences, Shanghai, 200030, P. R. China}

\author[0000-0002-8131-6730]{Yosuke Mizuno}
\affiliation{Tsung-Dao Lee Institute, Shanghai Jiao-Tong University, Shanghai, 201210, P. R. China}
\affiliation{School of Physics \& Astronomy, Shanghai Jiao-Tong University, Shanghai, 200240, P. R. China}
\affiliation{Key Laboratory for Particle Physics, Astrophysics and Cosmology (MOE), Shanghai Key Laboratory for Particle Physics and Cosmology, Shanghai Jiao-Tong University, Shanghai, 200240, P. R. China}

\author[0000-0003-4509-9705]{Bin Chen}
\affiliation{Institute of Fundamental Physics and Quantum Technology, \& School of Physical Science and Technology, \\Ningbo University, Ningbo, Zhejiang 315211, P. R. China}
\affiliation{School of Physics, \& Center for High Energy Physics,  Peking University, No.5 Yiheyuan Rd, Beijing 100871, P.R. China}

\begin{abstract}
We develop a semi-analytic framework for stationary, axisymmetric GRMHD jets that efficiently generates resolved polarized images from the near-horizon region out to $\sim 10^5\,r_g$ across a broad parameter space, enabling rapid exploration of how gravity and magnetohydrodynamic flows imprint scale-dependent signatures on jet morphology and polarization.
We identify a new scale-dependent separation in polarimetric diagnostics. Outside the photon ring, plasma loading strongly modifies the polarization-angle profile of the integrated jet-layer emission through inertia-driven winding of the magnetic field. At large image-plane radii, the polarization angle follows a power-law in radius, with an index determined by the jet collimation profile. Near the horizon, in contrast, jets converge to a universal polarization pattern controlled solely by black hole spin. This convergence is hierarchical: differences in velocity and magnetic-field structure are erased first, whereas collimation-dependent differences persist to smaller radii, thereby allowing these effects to be disentangled. These results establish a largely achromatic polarimetric diagnostic that connects GRMHD jet dynamics to resolved image structure, with direct implications for high-resolution polarimetry and for constraining black hole spin and jet formation.
\end{abstract}


\section{Introduction}
Relativistic jets are a common outcome of accretion onto spinning black holes and are thought to originate near the event horizon \citep{Blandford:1977ds,Blandford:1982xxl,Blandford:2018iot}. 
VLBI observations, particularly of M87*, now probe the environment of the black hole and jet-launching region directly \citep{hada2011origin,Doeleman:2012zc,Kim:2018hul,CraigWalker:2018vam,EventHorizonTelescope:2019dse,EventHorizonTelescope:2021bee,Lu:2023bbn,EventHorizonTelescope:2025uqi}. 
With upcoming high-resolution polarimetric measurements \citep{Park:2022vzb,Ayzenberg:2023hfw,Johnson:2023ynn,Johnson:2024ttr}, synchrotron polarization is becoming a powerful probe of jet dynamics and black hole spacetime \citep{ALMA:2025wvr}.

Deciphering the diagnostic potential of these observations requires a GRMHD description of magnetized plasma in strong gravity.
Although jet power is often discussed in terms of electromagnetic extraction from the ergoregion or accretion \citep{Blandford:1977ds,Blandford:1982xxl}, the interplay between matter and magnetic field, acceleration and collimation of baryon-loaded jets depend crucially on the plasma inertia \citep{Nitta:1991ui,Tomimatsu:2003uz,McKinney:2006dx,Tchekhovskoy:2011zx,McKinney:2012vh}. 
A key regime indicator is the degree of mass loading, inversely related to the magnetization and Lorentz factor, governing the efficiency of converting electromagnetic energy into kinetic energy as well as the degree of magnetic-field winding \citep{michel1969relativistic}. These quantities should be especially important in the bright millimeter jet sheath, where inertia is essential \citep{junor1999formation,McKinney:2006dx,Kovalev:2007hy,Moscibrodzka:2015pda, Cruz-Osorio:2021cob, Davelaar:2023dhl, Yang:2024kpz, Tsunetoe:2024eew}, unlike the highly magnetized, nearly force-free spine \citep{Gralla:2014yja,Chael:2023pwp}. 
Yet it remains unclear how image morphology responds to jet dynamics across different loading regimes, and which signatures provide robust diagnostics of plasma physics versus spacetime geometry.

A framework for inferring jet structures from polarization morphologies should be both physically transparent and computationally efficient. 
Numerical simulations are indispensable for time-dependent jet formation, but they are less suited to isolating individual physical effects or surveying broad parameter spaces. 
This motivates semi-analytic jet models that retain physical interpretability while enabling efficient parameter exploration. 
A number of phenomenological and force-free jet imaging models have been developed in this spirit \citep{Anantua:2019bna, Tsunetoe:2020pyz, Emami:2021ick, Tsunetoe:2022ktx, Papoutsis:2022kzp, Chael:2023pwp, Hou:2023bep, Zhang:2024lsf, Tsunetoe:2024uzh, Gelles:2024tpz, Tsunetoe:2025crz, Gelles:2026mxg, Jones:2026fbg}. 
However, no framework is both strictly GRMHD-based and flexible enough to link polarization directly to the flow while consistently capturing plasma and gravitational effects, hindering unified, multi-scale diagnostics.

In this Letter, we develop a semi-analytic framework for stationary, axisymmetric GRMHD jets that efficiently generates polarized images from the near-horizon region to very large radii. Our model links plasma loading, flow acceleration, and magnetic-field geometry to polarization structure, while remaining efficient enough for broad exploration of black hole, jet, and viewing geometries. Using this framework, we identify a scale-dependent separation of polarimetric diagnostics: in the acceleration region, polarization is strongly shaped by inertia-driven magnetic-field winding associated with plasma loading, whereas near the horizon, it converges to a universal pattern controlled solely by black hole spin. These results establish a novel polarimetric diagnostic connecting GRMHD jet dynamics to resolved image structure, with direct relevance for next-generation VLBI polarimetry.

\section{Jet Modeling}

We briefly outline the key equations underlying the jet model in this section. Given suitable boundary conditions, the plasma density, flow dynamics, and magnetic-field geometry can be determined self-consistently.
Throughout, we adopt $G=c=1$.

\subsection{Conservation laws in ideal GRMHD}

Although GRMHD is intrinsically complex, the conserved laws in stationary, axisymmetric ideal MHD provide the basis for a semi-analytical treatment \citep{bekenstein1978new}. We consider a magnetized baryonic plasma surrounding a stationary, axisymmetric black hole, with the spacetime line element
\bea
\mathrm{d}s^2 = g_{tt}\mathrm{d}t^2 + 2g_{t\phi}\mathrm{d}t\mathrm{d}\phi + g_{\phi \phi}\mathrm{d}\phi^2 +
g_{\mP\mP}\mathrm{d}\mP^2 \,,
\eea
where $\mP$ denotes the poloidal coordinates $(r,\t)$.
In the ideal-GRMHD limit of high conductivity, the magnetic field $B^{\m} = -(*F)^{t\m}$ is frozen into the poloidal flow, satisfying $\tilde{\epsilon}_{\mP\mP'} B^{\mP} u^{\mP'} = 0$, while the electric field $E^{\m} = F^{t \m}$ is induced by field-line rotation and frame dragging\footnote{These are the GRMHD primitive fields and are not covariant, but are easily related to physical fields. In the plasma comoving frame, $b^{t} = B^{\m}u_{\m}$ and $u^t b^i = B^i+ b^t u^i$. The fields measured by normal observers are $\mathcal{E}^{\m} = \a E^{\m}$ and $\mathcal{B}^{\m} = \a B^{\m}$, where $\a$ is the lapse function \citep{komissarov2004electrodynamics}.},
\bea\label{Bpsi}
\bag
& B^{\mP} = \tilde{\epsilon}_{\mP\mP'}\f{\partial_{\mP'}\psi}{\sqrt{-g}}\,, \quad B^{\p} =  (u^{\p} - \Omega_F u^t)\f{B^\mP}{u^\mP} \,, \\
& E^{\mP} = - g^{tt} g^{\mP\mP}\sqrt{-g} \left(\Omega_F - \omega \right) \tilde{\epsilon}_{\mP\mP'} B^{\mP'} \,,
\eag
\eea
and $E^{\p} = 0$, where $\tilde{\epsilon}_{\mP\mP'}$ denotes the Levi-Civita symbol, $\omega = -g_{t\p}/g_{\p\p}$ is the frame-dragging angular velocity, $\psi = A_{\p}$ is the stream function conserved along each field line and labeling magnetic flux surfaces, and $\Omega_F  = F_{t\mP}/F_{\mP\p}$ is the conserved field-line angular velocity \citep{Blandford:1977ds, thorne1982electrodynamics}. 
Notably, three magnetofluid quantities are conserved along field lines \citep{bekenstein1978new}: the mass flux $\eta$, the specific energy and angular momentum $\mE$, $\mL$, given by
\be
\bag
\label{conservedquantities}
&\eta  = \f{\rho\, u^{\mP} }{ B^{\mP}}, \,\,\,  \mE  = -h u_{t} - \f{\kappa B^{\phi} \Omega_F}{\eta} , \\
& \mL = \, h u_{\phi} -  \f{\kappa B^{\phi}}{\eta} ,
\eag
\ee
where $\rho$, $h$ are the ion density and specific enthalpy, $\kappa = g_{t\phi}^2 - g_{tt} g_{\phi\phi}$.
The conserved quantities $\{\psi, \Omega_F, \eta, \mE, \mL\}$ greatly simplify the flow solution. Combining Eqs.~\eqref{conservedquantities} with $u_\m u^\m = -1$ yields the relativistic wind equation \citep{camenzind1986hydromagnetic, takahashi1990magnetohydrodynamic}, which can be solved algebraically for the poloidal flow motion 
\bea
u_{p}^2+1 = h^{-2} \mathcal{F}_{\rm wind} \left(M_A^2, \mE,\mL, \Omega_F \right)\,,
\label{wind_eq}
\eea
where $M_A  = \left(u_p|\eta|/B_p\right)^{1/2}$ is the Alfv\'{e}n Mach number, with $u_p$, $B_p$
denoting the magnitudes of the poloidal velocity and magnetic field.
The function $\mathcal{F}_{\rm wind}$ is shaped by the spacetime geometry, and its explicit form is given in Appendix~\ref{App:basis}.
Here we focus on magnetically driven flows and adopt the (commonly used) cold limit, $h \to 1$, when solving the jet dynamics. Eq.~\eqref{wind_eq} then reduces to a quartic in $u_p$ \citep{camenzind1986hydromagnetic}, whose real solution is selected from root structures under specific conserved quantities. Subsequently, $u_t$, $u_{\p}$ follow from Eq.~\eqref{conservedquantities}, and the ion density is determined by $\rho = |\eta|B_p/u_p$. To obtain physically admissible solutions, we further impose the regularity conditions at the critical points, where $u_p$ matches one of the GRMHD wave speeds. Full details are given in \citep{Song:2025mhj}.

\subsection{Jet launching and asymptotics}\label{sec:asy}

Near the horizon, the plasma flows inward, but at larger radii it reverses to form an outflowing jet, with a stagnation region in between where the poloidal velocity vanishes \citep{Pu:2017akw,McKinney:2006tf,Huang:2019wqv,Chantry:2022ejm}. This transition is modeled as a geometrically thin stagnation surface (SS) \citep{camenzind1986hydromagnetic,takahashi1990magnetohydrodynamic}, which coincides with a local maximum of $k_0$ in the cold limit, with $k_0^{\prime} \lessgtr 0$ for the outflow and inflow branches.
Regularity of $B^{\phi}$ (Eqs.~\eqref{Bpsi}) at the SS enforces corotation, yielding 
$\Omega_F \mL  = \mE - \sqrt{k_0}\,\big|_{\rm SS}$.
Accounting for matter injection within the SS establishes a matching condition between the inflow and outflow \citep{Huang:2019wqv},
\bea
\label{eq:matching}
\f{ \mathrm{r}_{\eta} \,\mL_{\rm out} - \mL_{\rm in}}{ \mathrm{r}_{\eta} \,\mE_{\rm out} - \mE_{\rm in}}  = -\f{u_\p}{u_t} \bigg |_{\rm SS}\,, \quad  \mathrm{r}_{\eta} = \f{\eta_{\rm out}}{\eta_{\rm in}} \,, \label{matchBZ}
\eea
where ``in'' and ``out'' label the inflow and outflow, respectively. 
Another boundary condition is imposed at infinity, where we adopt a Michel-type outflow \citep{michel1969relativistic,goldreich1970stellar}. In this case, the fast magnetosonic (FM) point lies at infinity, providing a conservative estimate of the acceleration efficiency \citep{Gelles:2024tpz}. We then derive
\bea\label{eq:infity}
\mE_{\rm out} = \g_{\infty}^3\,, \quad \eta_{\rm out} = \f{\Omega_F^2 }{(\g_{\infty}^2 -1)^{3/2}}\left(B_pR^2\right) \bigg|_{r \rightarrow \infty} \,.
\eea
Here $(R,z) = r (\sin\theta, \cos\theta)$ denote cylindrical coordinates, and $\g_{\infty} = u^t|_{r \rightarrow \infty}$ is the asymptotic Lorentz factor, typically inferred from total-intensity observations \citep{biretta1999hubble,Hovatta:2008mj,Pushkarev:2017fbk}. In this work, $\g_{\infty}$ uniquely characterizes the magnetization and plasma loading. A very large $\gamma_\infty$ implies negligible $\eta_{\rm out}$, which is characteristic of force-free jets.

The inflow crosses an FM point located near the black hole, leading to more intricate acceleration behavior \citep{takahashi1990magnetohydrodynamic}. To determine the inflow solution, we combine: (i) the wind equation (Eq.\eqref{wind_eq}); (ii) the corotation condition at the SS; (iii) the matching conditions at the SS (Eqs.~\eqref{matchBZ}); and (iv) the regularity conditions at the FM point imposed by the acceleration equation.

\subsection{Field-line windings}\label{sec:winding}

Hereafter, we focus on the Kerr spacetime and work in Boyer-Lindquist (BL) coordinates. The stream function of jet is specified as $\psi(r, \theta) = C r^p\left(1- |\cos{\theta}|\right)$, favored by radio observations \citep{Asada:2011dr,Hada:2013yla,Pushkarev:2017fbk,Park:2020mbb,Burd:2021sjw} and numerical simulations \citep{Tchekhovskoy:2008gq,McKinney:2006dx}, where $0 < p < 2$ sets the degree of magnetic collimation\footnote{Although the exact form of $\psi(r,\t)$ near the black hole is determined by the trans-field equation \citep{fendt1995collimation,fendt1996collimated}, GR effects mainly modify the field only in the immediate vicinity of the event horizon \citep{Huang:2019wqv, Huang:2020lvl}. We therefore adopt a flat-spacetime approximation \citep{Broderick:2008qf}. The constant \(C\) is associated with GRMHD rescaling \citep{Gammie:2003rj}, chosen to ensure a consistent field strength at the SS.}. 
We focus on the field line anchored at the equatorial horizon $(r,\t) = (r_+,\pi/2)$, tracing the thin bright layer that dominates the jet morphology and is susceptible to plasma instabilities that can efficiently accelerate electrons and enhance nonthermal emission \citep{Dihingia:2021ncv,Davelaar:2023dhl,Yang:2024kpz}. The field-line angular velocity is then fixed by the regularity condition at the horizon \citep{Blandford:1977ds, znajek1977black}, which yields $\Omega_F = \Omega_H\left(3 - 2aM\Omega_H \right)^{-1}$ under magnetic-driven limit \citep{Gelles:2024tpz, Song:2025mhj}, where $a$ is the dimensionless spin parameter and $\Omega_H = aM(2 r_+)^{-1}$ is the black-hole angular velocity. 
\begin{figure}[htbp]
\centering
	\includegraphics[width=1\columnwidth]{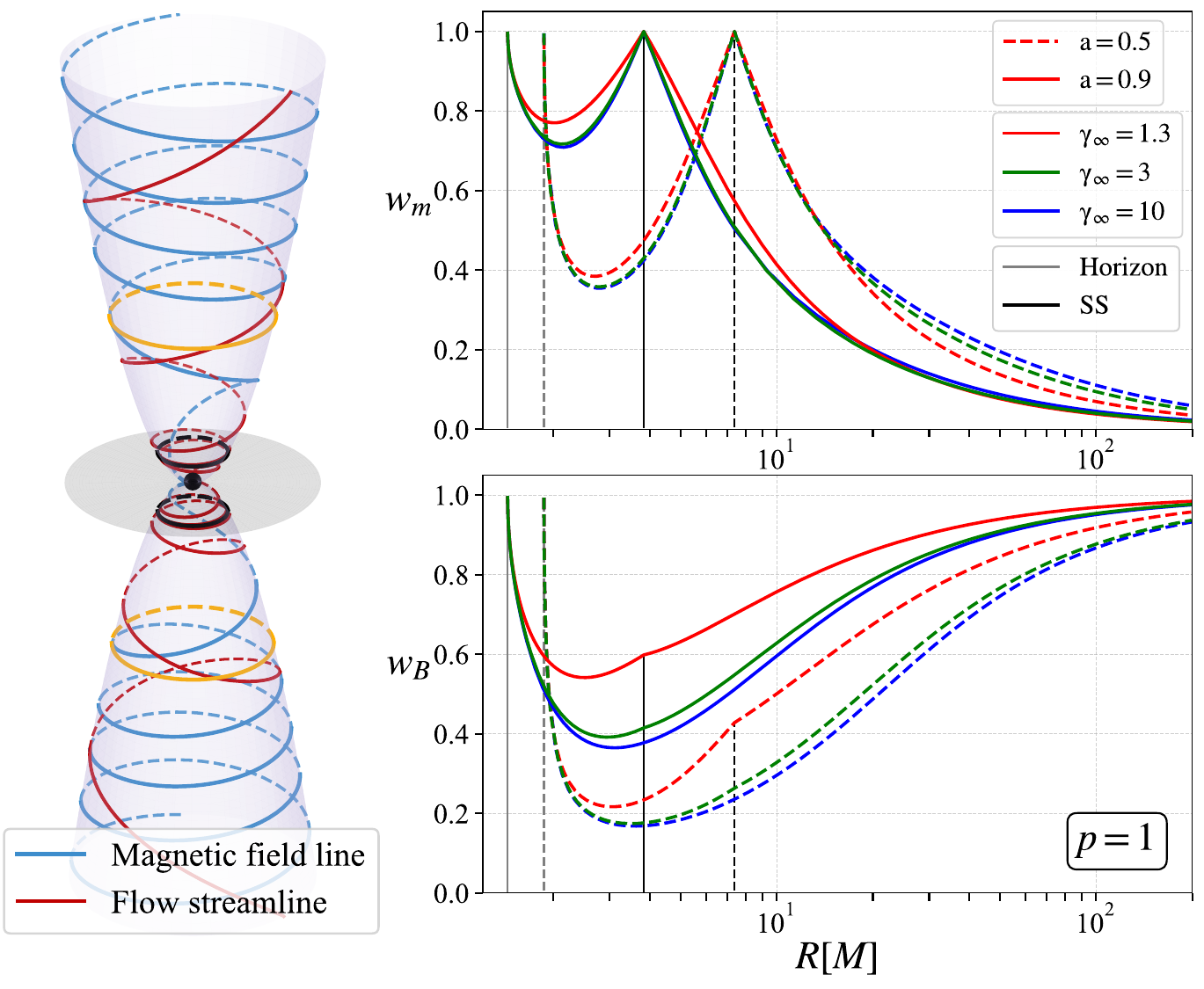}
	\caption{\textbf{Left:} Typical magnetic field line and flow streamline on the jet layer defined by $r_+ = r- |z|$, with $a = 0.5$, $\g_{\infty} = 10$. The black and yellow circles represent the SSs and outer light cylinders, respectively.
\textbf{Right:} Degree of streamline and magnetic field line windings on the jet layer.
The black and gray lines mark the SS and the event horizon, respectively. In the BL coordinates $u^{\p},B^{\p}$ formally diverge at the event horizon, leading to $w_m,w_B \to 1$. }
	\label{fig:winding}
\end{figure}

To illustrate how the flow streamlines and magnetic field lines evolve, we show in Fig.~\ref{fig:winding} the flow and magnetic-field winding degrees, defined as
\bea
w_m = \f{2}{\pi}\tan^{-1}\left(\f{u_T}{u_p}\right), \quad 
w_B = \f{2}{\pi}\tan^{-1}\left(\f{B_T}{B_p}\right),
\eea
where $u_T = \sqrt{g_{\p\p}}\,|u^{\p}|$ and $B_T = \sqrt{g_{\p\p}}\,|B^{\p}|$ denote the toroidal components. 
Along the outflow branch, once the plasma crosses the light cylinder, the magnetic field lines become increasingly wound; the flow is progressively accelerated in the poloidal direction, driving $w_m \to 0$. A higher black-hole spin also produces stronger magnetic winding.

Clearly, a lower $\g_{\infty}$, meaning heavier plasma loading, leads to more wound magnetic fields.
As $\g_{\infty}$ decreases from $10$ to $1.3$, $w_B$ increases by $40\%$ at $x \sim 10 M$. 
At large distances, as $B_T \rightarrow 2\psi\Omega_F \gamma_{\infty}\left(\gamma_{\infty}^2-1\right)^{-1/2}$ and $B_{p} \sim \mO(r^{p-2})$, the magnetic field becomes predominantly toroidal. By contrast, the flow winding ($w_m$) is less sensitive to $\g_\infty$.
This trend is essential for imaging analysis: plasma inertia modifies both the acceleration efficiency and the degree of field-line twisting, thereby leaving distinct imprints on jet polarimetry.

\section{Polarization Morphology}

To efficiently explore polarized images, we develop a semi-analytical general relativistic radiative transfer (GRRT) scheme for the jet model, in which the Stokes parameters are accumulated linearly along each light ray at its intersections with the emission layer.
We adopt a widely used nonthermal electron population with a power-law electron distribution function (eDF) \citep{EventHorizonTelescope:2021btj, Gelles:2024tpz} to compute the local synchrotron emissivity, and further modulate it with a beam-like function to capture eDF anisotropy \citep{Lai:2025yeq, Zhou:2025moa}.
Details of the emission model and radiative transfer are given in Appendix~\ref{App:SRT}.
Combining the GRMHD conservation laws, critical-point analysis, and GRRT, we generate polarized jet images over a range of spins, jet parameters, and emission and viewing geometries.

\begin{figure}[htbp]
	\includegraphics[width=1\columnwidth]{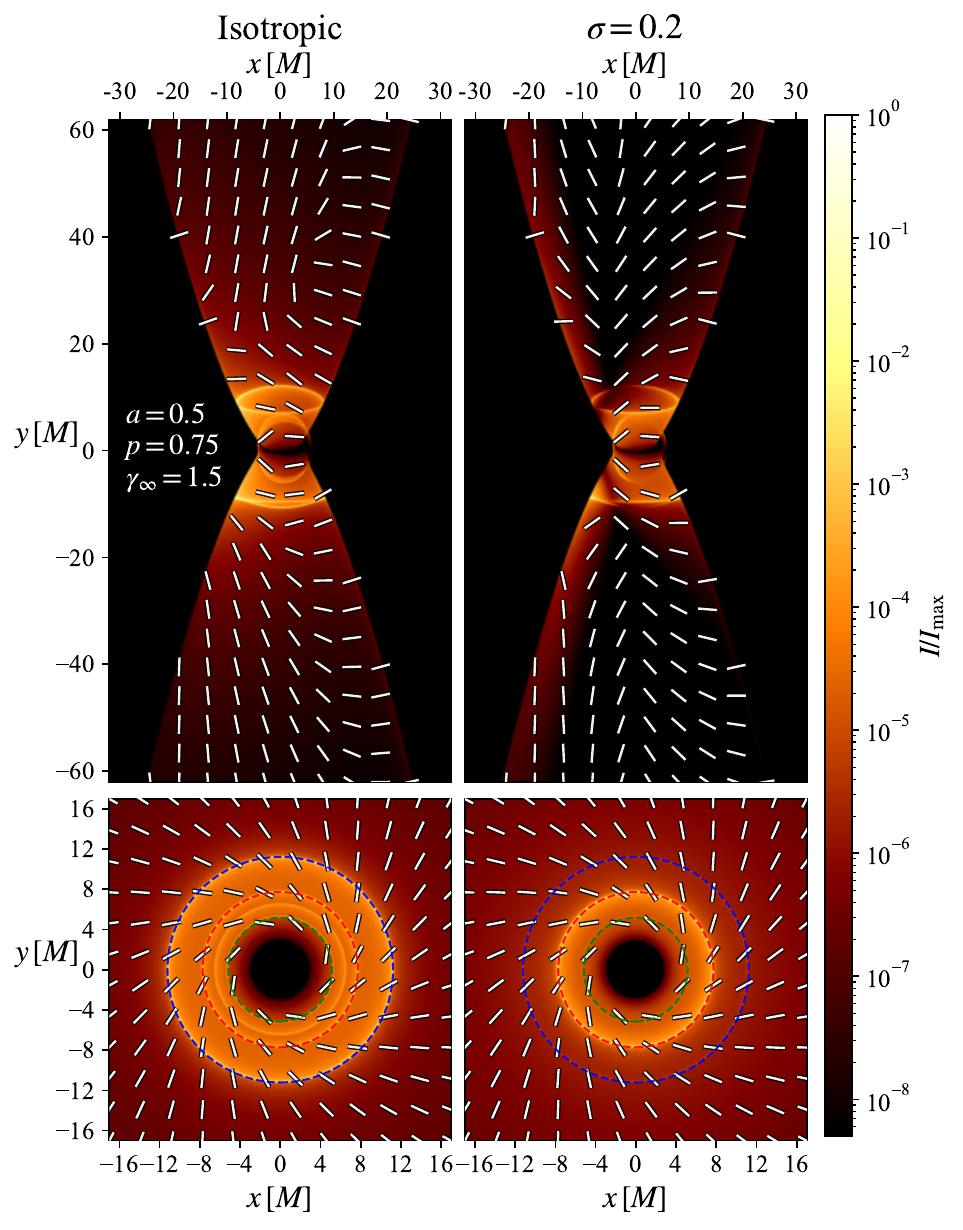}
	\caption{Intensity maps overlaid with EVPA of the jet layer, computed for $a = 0.5$, $p = 0.75$ and $\g_{\infty} = 1.5$. 
The left panels correspond to an isotropic eDF, while the right panels show results for a bi-beam-like eDF with $\sigma = 0.2$. The top and bottom rows represent a nearly edge-on observer ($\t_o = 80^{\circ}$) and a face-on observer ($\t_o = 0.01^{\circ}$). For the face-on view, the dashed blue and red curves show the direct images of the SSs on the counter-jet and forward-jet sides, respectively, while the green curve indicates the photon ring.}
	\label{fig:jetimage}
\end{figure}

Fig.~\ref{fig:jetimage} shows the visualization of the jet images for nearly edge-on and face-on views. 
In the edge-on case, a bright jet base appears near the black hole \citep{Kawashima:2020rmr}, associated with dense inflow \citep{Song:2025mhj}, with the SS forms a bright, oblate ring \footnote{The density formally diverges at the geometrically thin SSs, which acts as the only source of plasma loading. In this work, we cap the density at three times its horizon value and smooth it with a cubic spline (see Fig.\ref{fig:density}).}. 
The polarization vectors predominantly align with the $y$-direction in the core, but rotate toward the edges, becoming largely perpendicular to the jet limb as a projection effect associated with helical magnetic fields \citep{Attridge:1999fw, pushkarev2005spine, Lyutikov:2004kn, Gabuzda:2014tza, Baghel:2023equ}. 
Due to lensing and aberration from the toroidal motion, the polarization vectors are asymmetric about the $y$-axis, resulting in a nonzero net polarization \citep{lyutikov2003polarization}.
In the face-on view, there is a multi-ring structure, composed of the forward-jet's SS, the photon ring, and the counter-jet's SS, from small to large.

The eDF anisotropy modulates the intensity map by altering the pitch-angle dependence of the local emissivity. In the edge-on view, this leads to limb brightening, as rays passing through the jet core and edge sample different pitch-angle distributions. However, the eDF anisotropy only slightly affects the EVPA. This implies that the polarization architecture is largely dictated by macroscopic GRMHD flow dynamics and spacetime geometry.
We therefore restrict our subsequent analysis to the isotropic case. The anisotropic eDF is discussed further in Appendix~\ref{App:aniso} and will be explored in future work \citep{jetaniso:2026}.

\subsection{Imprint of plasma loading}\label{sec:acceEVPA}

\begin{figure}[htbp]
      \includegraphics[width=1.02\columnwidth]{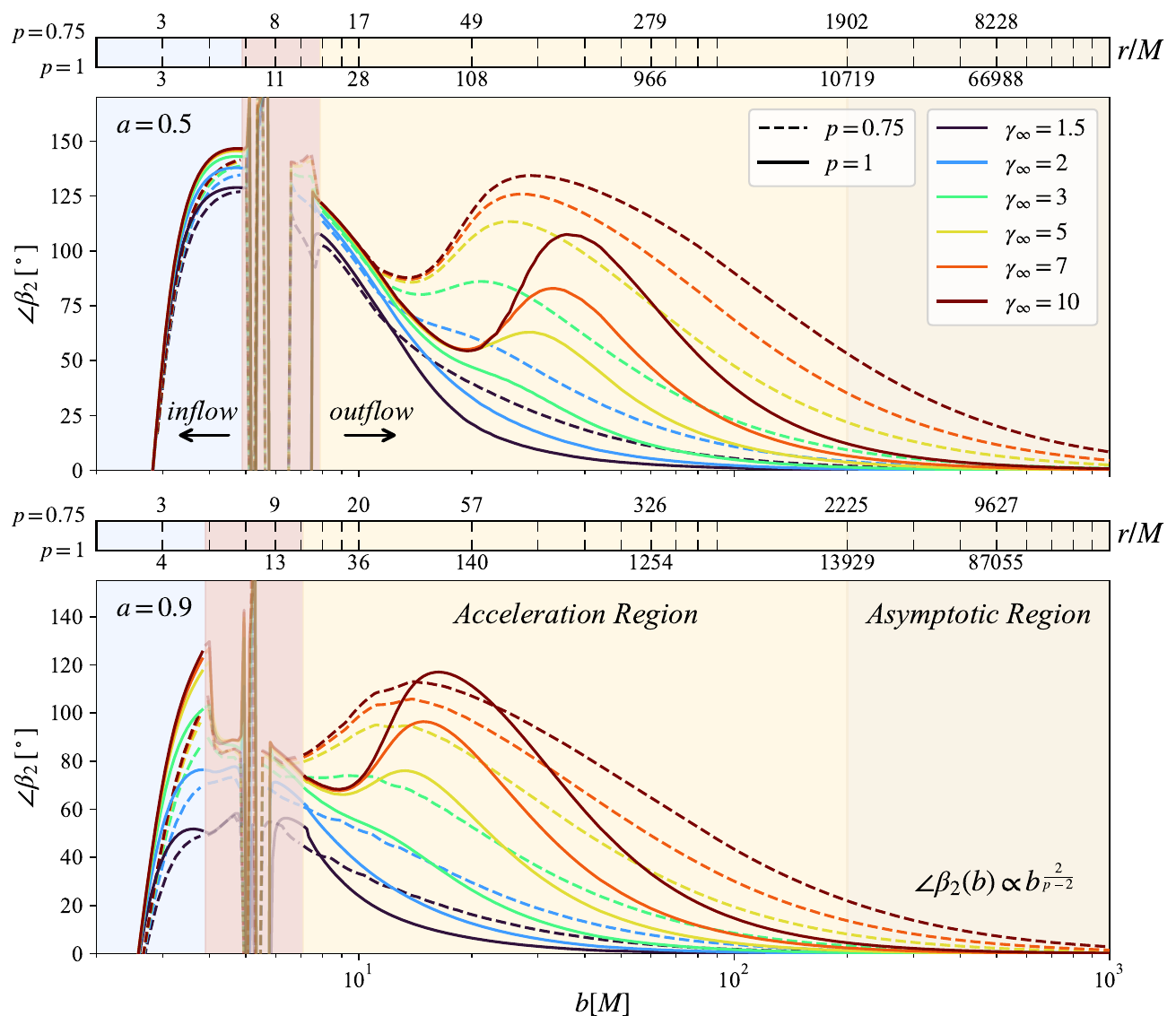}
	\caption{Polarization pattern vs. image-plane radius $b = \sqrt{x^2 + y^2}$ for a nearly face-on view, produced along the jet with various $\g_{\infty}$ and $p$. From left to right, the colored regions mark the near-horizon region, lensing band, acceleration region, and asymptotic region. In the asymptotic region, the GR effect is negligible for MHD flow, light emission, and propagation, and one can effectively work within the framework of special relativistic radiative transfer (SRRT) (Appendix~\ref{App:SRRT}).}
	\label{fig:windingeta2}
\end{figure}

To highlight the polarimetric trends, we focus on the nearly face-on case, which allows a transparent analysis and is relevant to a broad class of astrophysical systems \citep{o2009three, EventHorizonTelescope:2020dlu, Kovalev:2025kxf, Mertens:2016rhi}. As shown in Appendix~\ref{App:inclinaiton}, the main features remain robust at moderate viewing inclinations.
As a direction-independent diagnostic of the nearly axisymmetric electric vector position angle (EVPA), we use the phase of the second Fourier coefficient in the azimuthal decomposition of the linear Stokes parameters, $\arg(\b_2)$ \citep{Palumbo:2020flt}, with $\arg(\b_2) \simeq 2 \,\text{EVPA}\big|_{\phi = 0^\circ}$. Fig.~\ref{fig:windingeta2} shows $\arg(\b_2)$ as a function of image-plane radius $b$. We distinguish four regions with different underlying physics: the near-horizon region, the lensing band, the acceleration region, and the asymptotic region ($b > 200M$), indicated by different shaded colors. The narrow lensing band is produced by strongly lensed rays that cross the equatorial plane twice \citep{Gralla:2019xty,Johnson:2019ljv}, leading to sharp transitions in both intensity and EVPA.

In the acceleration region, the EVPA is sensitive to the plasma mass loading. 
It varies by up to $100^{\circ}$ between a moderately relativistic flow ($\g_{\infty} = 1.5$) and a highly relativistic one ($\g_{\infty} = 10$). 
Highly relativistic flows approach the force-free regime \citep{Gelles:2024tpz}, in which the EVPA exhibits a clear ``bump'' near $b \sim 50 M$, caused by combined emission from forward and counter jets: gravitational lensing enhances the counter-jet emissivity by increasing its pitch angle within the light cylinder. The leading edge of the bump is thus primarily associated with counter-jet emission. On larger scales ($r \gtrsim 10^3\,M$), the forward jet dominates as Doppler boosting becomes significant. The bump profile is further shaped by the collimation index and black-hole spin.

Plasma loading smooths the EVPA profile by reducing the pitch-angle effect, reflecting the close coupling between mass loading and field-line twisting.
A heavier loading (smaller $\gamma_{\infty}$) enhances the role of plasma inertia in the GRMHD flow and, in turn, leads to stronger magnetic-field winding (i.e., a larger $w_B$ in Fig.~\ref{fig:winding}).
The pitch-angle contrast between the forward and counter jets, mainly induced by the poloidal magnetic field, is then suppressed. In the weak-lensing regime ($b \gg M$), the pitch angles for a single ray crossing the counter and forward jets with 
\footnote{The same qualitative trend persists for general $p$, although the expressions become more complicated and are therefore not presented here.}
$p = 1$, denoted by $\alpha_{\rm c}$ and $\alpha_{\rm f}$, can be derived as
\bea\label{eq:paio}
\cos{\alpha_{\rm c}} \simeq  \f{\cos(4M b^{-1})}{\sqrt{1+w_{\rm c}^2}} \,, \quad
\cos{\alpha_{\rm f}} \simeq  \f{\cos(r_+ b^{-1})}{\sqrt{1+w_{\rm f}^2}} \,,
\eea
where $w_{\rm f}$ and $w_{\rm c}$ denote the ratios of $B_T$ to $B_p$ at the crossing points.
For small winding, $\alpha_{\rm c} \simeq 4M b^{-1} > \alpha_{\rm f} \simeq r_+ b^{-1}$, and the counter-jet emission can dominate in the acceleration region.
As $\g_{\infty}$ decreases, both $w_{\rm f}$ and $w_{\rm c}$ increase, thereby suppressing the pitch-angle difference.
In the limit $B_T \gg B_p$, both $\alpha_{\rm c}$ and $\alpha_{\rm f}$ approach $\pi/2$, and the EVPA becomes dominated by the Doppler boosting forward jet.
As in Fig.~\ref{fig:windingeta2}, for $\gamma_{\infty} \leq 3$, the bump feature nearly disappears and the EVPA profile is smoother.
If the eDF anisotropy is taken into account, a larger pitch angle could instead lead to lower emissivity, and the counter jet emission makes a smaller contribution, and the bump will be smoother.

\subsection{Near-horizon convergence}\label{sec:NHP}

\begin{figure}[htbp]
\centering
 \includegraphics[width=1\columnwidth]{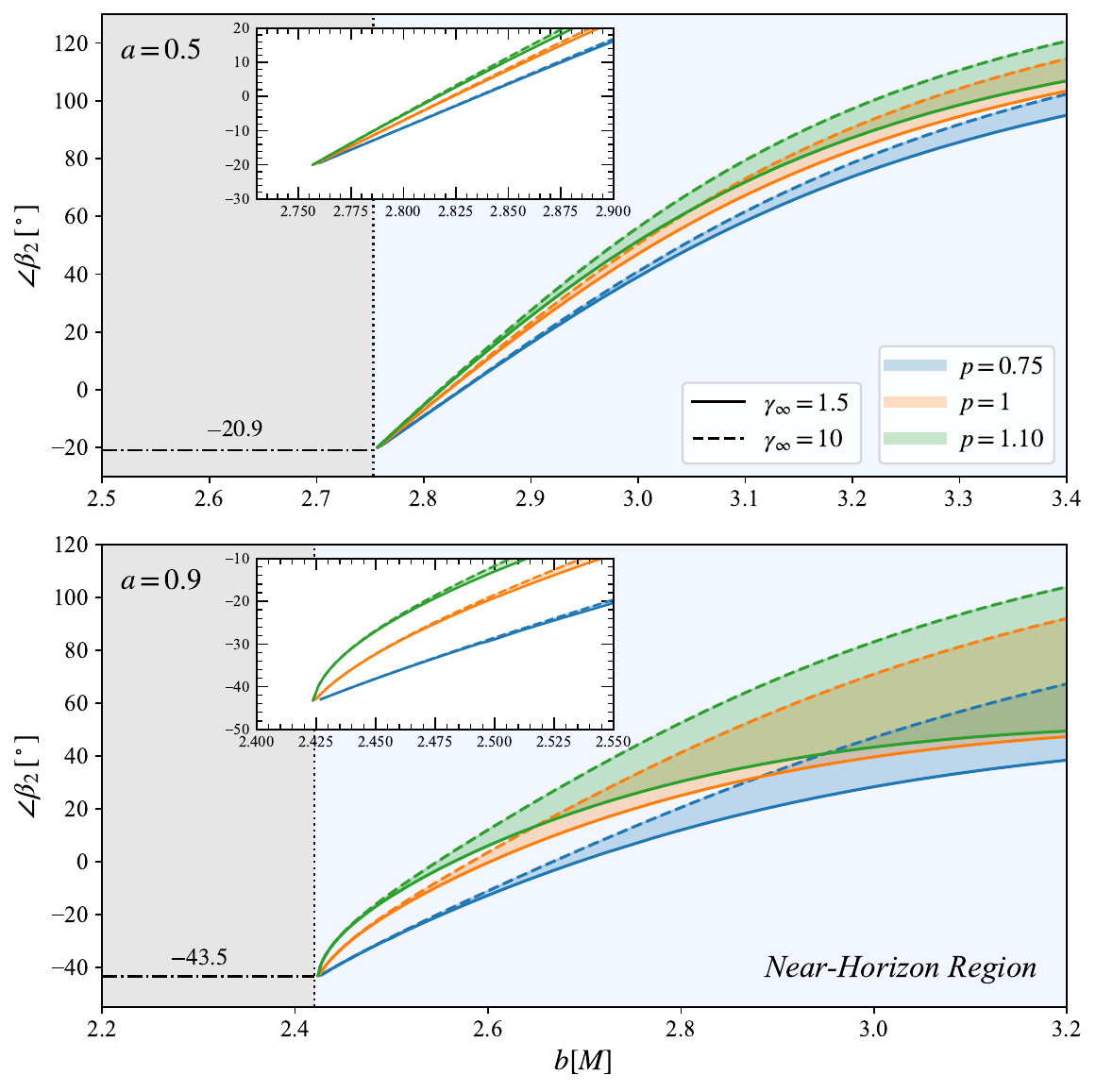}
	\caption{Near-horizon EVPA of the jet layer anchored to the equatorial horizon. Results for different collimation indices are shown in different colors. Within each colored region, the asymptotic Lorentz factor increases from 1.5 (solid curve) to 10 (dashed curve) from bottom to top.}
	\label{fig:NHPjet}
\end{figure}

In the near-horizon region, frame dragging greatly twists the inflow and magnetic field into toroidal patterns, leading to a radial EVPA pattern.   
When getting extremely close to $r_+$, different flows are dragged to degenerate to a unified pattern, causing $\arg(\b_2)$ for different $\g_{\infty}$ and $p$ converge to a universal near-horizon polarization (NHP) pattern, as shown in Fig.~\ref{fig:NHPjet}. 
This universality was first shown in our equatorial disk model \citep{Hou:2024qqo} and later extended to more general cases \citep{Chen:2025ysv, Chael:2026fhf, Hou:2026snq}.
Our jet results further extend this picture. In particular, we derive the following formula:
\bea\label{eq:nhp}
\begin{aligned}
\arg(\b_2) &=  -2\tan^{-1}\Big[ \f{a}{\sqrt{b_+^2-a^2}} \Big] \\
& + \D \,\f{\Omega_H^2}{a^2}\,\f{\sqrt{b_+^2-a^2} \left( 1-a\,\Omega_F \right) + p \,r_+}{\Omega_H - \Omega_F} + \mathcal{O} (\D^2)\,,
\end{aligned}
\eea
where $\D = r^2-2Mr+a^2$ measures the distance to the horizon, and $b_+$ corresponds to the projected horizon radius. Clearly, the leading-order term depends only on the spin, whereas the next-to-leading-order term also encodes the effects of $p$ and $\Omega_F$. This suggests that the jet shape and field-line rotation have a stronger influence on the NHP than the flow velocity and magnetization degree. 
As verified in Fig.~\ref{fig:NHPjet}, as the flow approaches $r_+$, models with different $\g_\infty$ rapidly converge to a common EVPA, whereas those with different $p$ do so more slowly. Therefore, the near-horizon trend can help disentangle the effects of $\g_\infty$ and $p$.

For general viewing angles, variations in the incidence angle of light rays introduce an asymmetric NHP distribution along the jet base, but do not affect its leading-order flow independence \citep{Hou:2024qqo}. Changes in the eDF, jet poloidal geometry, and toroidal field-line rotation do not modify the leading-order NHP, which is shaped by irresistible frame dragging.
Combined with previous results, this implies that the horizon EVPA robustly probes the spacetime geometry, regardless of whether the emission is disk-dominated or jet-dominated, since the accumulated polarization is governed primarily by photon trajectories rather than by plasma effects.

\subsection{Asymptotic structure}\label{sec:infEVPA}

In the asymptotic region well outside the light cylinder, the EVPA is generally governed by the MHD dynamics itself. 
As the plasma loading alters both the azimuthal magnetic field and the flow velocity \citep{Nitta:1991ui, takahashi1998trans}, it essentially modifies the asymptotic EVPA there. 
Evaluating the flow velocity and magnetic field in the large-$r$ limit yields the asymptotic transverse polarization vector, $f^{\t} \sim \mO(1) \gg f^{\p} \sim \mO(r^{p/2-1})$, for any $\gamma_{\infty}$ and $0 < p < 2$, which fully determine the EVPA.
The asymptotic polarization for the MHD flow then takes
\bea\label{eq:evpamhd}
\arg(\b_2)
 &\to&
\f{p-2}
{\f{\g_{\infty}}{\sqrt{\g_{\infty}^2-1}}-1}\, \f{1}{\Omega_F r} \,. 
\eea
Hence, $\arg(\b_2)$ decays as $r^{-1}$ and converges to a purely radial pattern, a universal trend seen in Fig.~\ref{fig:windingeta2}. This arises from the asymptotically toroidal magnetic-field configuration (Fig.~\ref{fig:winding}), which dominates over the plasma's aberration effect and drives this trend. Combined with the asymptotic jet-shape scaling $b \to R \simeq \sqrt{2\psi}\, r^{1-p/2}$, Eq.~\eqref{eq:evpamhd} implies a power-law relation
\bea
\arg{\beta_2}(b) \propto b^{\f{2}{p-2}} \,,
\eea
with the coefficient set by $\g_\infty$ and $\Omega_F$. This provides a new, observation-relevant constraint on jet collimation and related parameters.

In the idealized force-free electrodynamics (FFE) regime, the jet is completely devoid of matter inertia, and its asymptotic EVPA retains a geometry-dependent value governed by the collimation index $p$ \citep{Gelles:2024tpz}. However, this cannot be achieved by taking $\g_\infty\to +\infty$ in Eq.~\eqref{eq:evpamhd}. 
This uncovers a fundamental non-commutativity of two limits: FFE and large $r$. 
For any MHD flow with a finite mass-loading degree, plasma inertia inevitably drives the magnetic field into an asymptotically toroidal configuration at infinity, forcing the EVPA to universally decay as $r^{-1}$ regardless of the choice of $p$ (see Appendix~\ref{App:infevpa}). 
This qualitative discontinuity implies that FFE cannot be continuously connected to baryonic outflows at large distances, emphasizing the absolute necessity of a full GRMHD treatment when tracking the EVPA from near the horizon to infinity.

\section{Summary and Discussion}

We have developed a semi-analytic framework that connects stationary, axisymmetric ideal GRMHD jet dynamics to polarimetric observables across scales, from the event horizon to spatial infinity. By varying the plasma mass loading ($\g_{\infty}$), collimation ($p$), and spin, we find that the scale-dependent EVPA morphology encodes distinct physical information in three regimes.

\textbf{Acceleration region.} The EVPA profile is controlled by the interplay between flow acceleration and field-line twisting. 
In strongly magnetically dominated flows, lensed counter-jet emission threaded by poloidal field produces an EVPA bump, whereas larger baryon loading strengthens the toroidal field and suppresses this feature, including at moderate inclinations. 
Polarimetric imaging of this region can therefore help constrain the jet matter content and mass-loading efficiency, particularly when combined with total-intensity constraints on the jet Lorentz factor \citep{Homan:2014uea,Pushkarev:2017fbk,Kutkin:2018qmn,Kino:2022xme}. This transition should become accessible to next-generation EHT and space-VLBI measurements \citep{Johnson:2023ynn, Johnson:2024ttr}.

\textbf{Near-horizon gateway.} Frame dragging drives different inflow solutions toward a common NHP pattern. The imprint of $\g_\infty$ is rapidly erased as the flow approaches the horizon, while the dependence on $p$ survives to smaller radii before also converging. 
The NHP is therefore largely insensitive to jet microphysics and may provide a comparatively clean probe of black hole spin. This regime is well matched to the angular resolution expected from future space-based interferometric concepts \citep{Johnson:2019ljv}.

\textbf{Asymptotic far field.} For nearly face-on viewing, the MHD flow exhibits a universal asymptotic $r^{-1}$ EVPA decay (Eq.~\eqref{eq:evpamhd}). The derivation exposes a non-commutativity between the force-free and large-$r$ limits and predicts a power-law relation between EVPA and image-plane radius, with the index set by the collimation. This behavior provides a potentially testable signature on parsec scales, for example, with the VLBA \citep{napier1994very}.

Taken together, these results show that multi-scale EVPA structure can be used to disentangle plasma loading, field-line twisting, jet collimation, and black hole spacetime in a unified framework. In particular, our analysis identifies three robust signatures: the dependence of the acceleration-region EVPA bump on baryonic loading, the near-horizon convergence of polarization morphology, and the collimation-controlled asymptotic EVPA profile.

Several caveats should be kept in mind. 
The large-scale EVPA patterns discussed here are geometric in origin,
but time-dependent disturbances such as shocks, magnetic reconnection, and MHD instabilities can induce local, transient fluctuations superposed on the time-averaged EVPA structure \citep{marscher1985models,sironi2015relativistic,Mizuno:2012xd}. 
Faraday rotation and differential depolarization may additionally modify the observed polarization at longer wavelengths \citep{burn1966depolarization}, although the intrinsic morphology should remain more directly accessible at sufficiently high frequencies (e.g., $\gtrsim 345$ GHz), as indicated by EHT works \citep{EventHorizonTelescope:2021bee}. 

Our treatment also assumes single-fluid baryonic loading near the stagnation surface; more realistic plasma composition, including pair-rich spine--sheath structures, may alter both the dynamics and emissivity \citep{lightman1987pair,Blandford:1995yf,sikora2000pair,Levinson:2010fc,Broderick:2015swa}. In addition, eDF anisotropy and finite emission-region thickness may introduce further modifications through pitch-angle-dependent synchrotron emissivity and line-of-sight depolarization \citep{sokoloff1998depolarization}. Quantifying the robustness of the predicted EVPA morphology under these effects will be an important direction for future work.


\begin{acknowledgments}
The work is partly supported by NSFC Grant No. 12275004, 11735001, 12588101, 12547123, and 12547127. YM is supported by the National Key R\&D Program of China (grant No. 2023YFE0101200), the National Natural Science Foundation of China (grant Nos. 12273022 and 12511540053),  and the Shanghai Municipality Orientation Program of Basic Research for International Scientists (grant No. 22JC1410600).

\end{acknowledgments}

\newpage
\appendix

\section{Stationary, axisymmetric GRMHD flows}
\label{App:1}

\subsection{Basic setup}
\label{App:basis}

\begin{figure}[htbp]
\centering
      \includegraphics[width=3in]{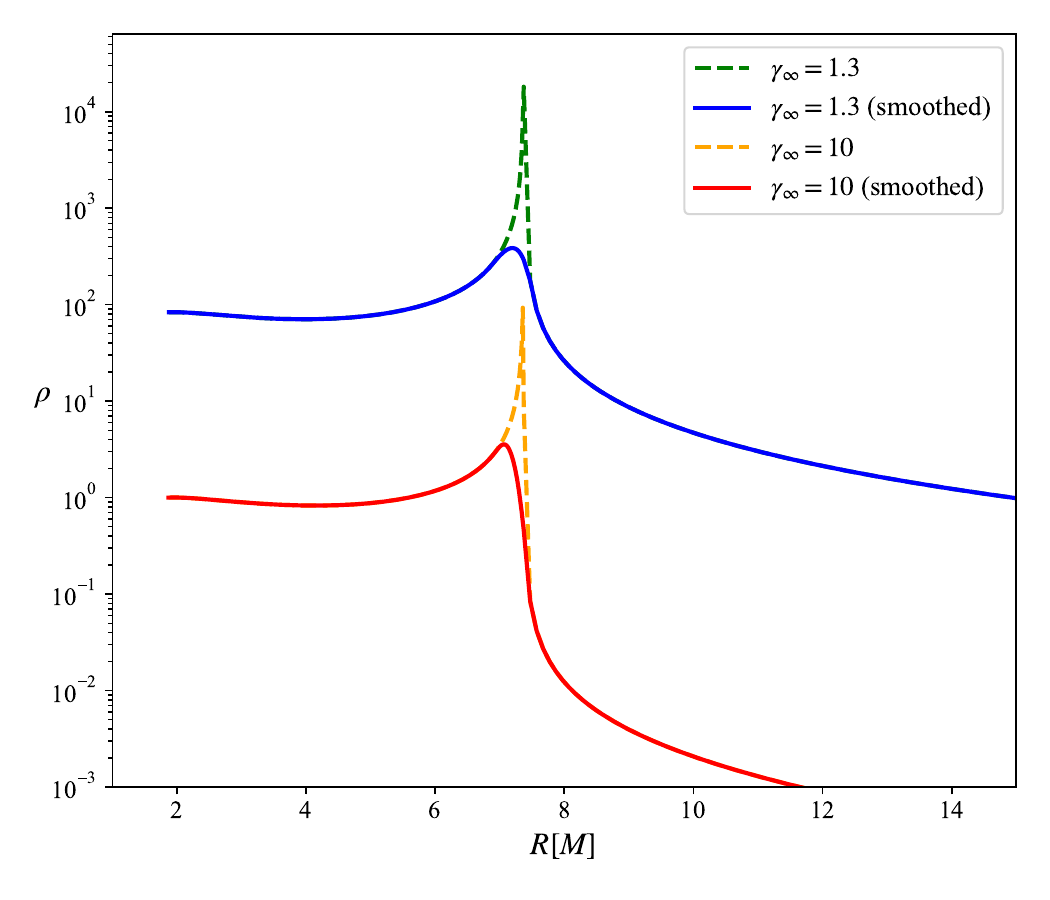}
	\caption{Plasma number density $\rho$ along the jet layer defined by $r_+^p = r^p(1- |\cos{\theta}|)$ under $a = 0.5$ and $\g_{\infty} = 1.3$ or $10$. Because we assume a geometrically thin stagnation surface, mass loading causes $\rho$ to formally diverge there, as indicated by the dotted curve. To remove this divergence, we truncate the density at three times its value on the horizon and apply a cubic spline interpolation to smooth the profile. The resulting smoothed distribution is shown by the solid curve for each parameter set.}
	\label{fig:density}
\end{figure}

In this section, we review the basic equations for the construction of the jet model, containing the expressions for the flow velocity, magnetic field and plasma density.
The temporal and azimuthal components of the four-velocity are determined by inverting the definitions of the conserved quantities $\mE,\mL$ in Eq.~\eqref{conservedquantities}. Explicitly, under the cold limit, $u_t,u_{\p}$ are given by
\be
\label{eq:utup}
\bag
u_t = \mE \f{(g_{tt}+g_{t\p}\Omega_F)(1 - \Omega_F l) + M_A^2}{k_0 - M_A^2} \,,  \quad 
u_\p = \mE \f{(g_{t\p}+g_{\p\p}\Omega_F)(1 - \Omega_F l)-M_A^2l}{k_0 - M_A^2} \, ,
\eag
\ee
where $k_0 = -\left(g_{\p\p}\Omega_F^2+2g_{t\p}\Omega_F+g_{tt}\right)$, $l = \mL/\mE$ is the angular momentum density of the flow. Substituting the above equations into $B^{\p}u^\mP =  (u^{\p} - \Omega_F u^t)B^\mP$ yields the explicit expression for the toroidal magnetic field, which is
\bea\label{eq:Bphi02}
B^{\phi}= \eta \mE \,\f{ (g_{tt} + g_{t\phi} \Omega_F) l + (g_{t\phi} + g_{\phi\phi} \Omega_F)  }{\kappa (k_0 - M_A^2) }\,.
\eea
Hence, while the poloidal magnetic field is obtained by specifying the stream function, the toroidal component is coupled to the flow and depends sensitively on the plasma loading.
In the zero-mass-loading limit, $\eta \to 0$ with $\eta \mE$ remaining finite, we recover the force-free result, $\k \Omega_F B^{\p} = - \eta \,\mE$.  
By combining  Eqs.~\eqref{eq:utup} with the normalization condition of the flow velocity, $u_\m u^\m = -1$, we can obtain the relativistic Bernoulli equation (also called the wind equation) that governs the evolution of the poloidal velocity along the field lines \citep{camenzind1986hydromagnetic, takahashi1990magnetohydrodynamic}.
Through the ideal MHD condition, we have $u^{\t} = B^{\t}u^r/B^r$, $u^r = \pm u_p \left(g_{rr}+g_{\t\t}(B^\t)^2/(B^r)^2\right)^{-1/2}$, where the sign ``$\pm$'' corresponds to outflow/inflow. 
The wind equation can then be written as 
\bea\label{eq:windapp}
\left(u_{p}^2+1\right)(k_0-M_A^2)^2 = \mE^2 \left( k_0k_2-2k_2M_A^2-k_4M_A^4\right)\,, \quad 
M_A^2  = \f{\eta u_p}{B_p} \,,
\eea
where the functions take
\bea
k_2= \left(1-\Omega_F l\right)^2\,, \quad
k_4= -\f{g_{\p\p}+2g_{t\p}l+g_{tt}l^2}{\k}\,.
\eea
The Alfv\'{e}nic Mach number $M_A$ characterizes the degree of magnetic acceleration of the flow.     
Equation~\eqref{eq:windapp} is a quartic equation in $u_p^2$, under the cold limit; for given $\{\eta,\mE,\mL\}$, it typically admits one physically relevant solution and three unphysical branches.
Regular jet solutions do not exist for arbitrary choices of the conserved quantities.
They must satisfy the requirement that the flow passes smoothly through all critical points  \citep{takahashi1990magnetohydrodynamic}, defined by $u_p$ matching the characteristic MHD wave speeds: the slow and fast magnetosonic speeds and the Alfv\'{e}n speed.    
As discussed in Sec.~\ref{sec:asy}, we obtain explicit solutions for $\eta_{\rm out}$, $\mE_{\rm out}$ for the outflow by adopting a minimal-energy Michel solution \citep{michel1969relativistic,goldreich1970stellar}, in which the fast magnetosonic (FM) point is pushed to infinity. In this case, the conserved quantities reduce to
\bea\label{eq:infity1}
\mE_{\rm out} = \g_{\infty}^3\,, \quad \eta_{\rm out} = \f{\Omega_F^2 }{(\g_{\infty}^2 -1)^{3/2}}\left(B_pR^2\right) \bigg|_{r \rightarrow \infty} \propto \f{\Omega_F^2 }{(\g_{\infty}^2 -1)^{3/2}} \,,
\eea
where $\g_{\infty} = u^t\big|_{r \rightarrow \infty}$  is the asymptotic Lorentz factor of the outflow; $R = r \sin{\theta}$ is the cylindrical radius. Since the jet is well collimated, $R$ serves as the natural radial coordinate across the flow.
The scaling $B_p \sim R^{-2}$ is a necessary condition for Michel-type solutions. This behavior is also realized by the asymptotic magnetic field associated with the stream function $\psi = r^p \left(1-|\cos{\t}|\right)$, which gives
\bea\label{eq:Bflat}
\begin{aligned}
&B_\mu \xrightarrow{r \to \infty} 
\left\{
0,\;
r^{p-2},\;
-\f{p\psi}{r\sin\theta},\;
r^2\sin^2\theta\,B^\phi
\right\} \\
&B^{\p}  \xrightarrow{r \to \infty} 
 -\f{\Omega_F \g_\infty B_p}{(\g_\infty^2 - 1)^{1/2}}
 \,,
\end{aligned}
\eea
from which we have $B_p \propto R^{-2}$ for any index $0<p<2$. 
Once the mass flux is determined and the poloidal velocity is solved through the wind equation (Eq.~\eqref{eq:windapp}), the plasma density is directly obtained through the relation
\bea
\rho = \f{B_p |\eta|}{u_p}
\eea
for both inflow and outflow. At the jet launching region, the poloidal velocity goes to zero, and $\rho$ formally diverges. 
This divergence reflects the idealized treatment of plasma injection at an infinitely thin stagnation surface (SS), which separates the inflow and outflow. It has been verified that the function $k_0$ acts as an effective potential controlling its location: the condition $ k_0^{\prime} \lessgtr 0$ for outflow/inflow implies that the SS is located at a local maximum of $k_0$  \citep{Song:2025mhj}.
Such an infinitesimally thin, dense loading layer would otherwise produce an extremely bright ring in synthetic jet images.
In this work, we truncate the density near the SS at three times its value on the horizon, and apply a cubic spline interpolation to smooth the profile. This can be understood as effectively replacing the infinitesimally thin SS with a finite-thickness transition layer, within which the detailed density structure is not resolved and the matching conditions (Eq.~\eqref{eq:matching}) are imposed on its two sides. The resulting smoothed distribution is shown in Fig.~\ref{fig:density}. From this figure, the inflow density is clearly higher than that of the outflow, arising from the black hole's gravitational pull on the matter launched from the SS. Although the outflow can, in certain regimes, be well described by force-free electrodynamics, the inflow generally cannot; therefore, a global GRMHD framework is required.

Near the event horizon, $r \to r_+$, the inflow solution admits a special asymptotic form.
In the BL coordinates, by expanding the wind equation Eq.~\eqref{eq:windapp} in terms of $\D=r^2-2r+a^2$ and using the expression for the stream function $\psi = r^p \left(1 - |\cos{\t}|\right)$, we get the poloidal velocity as
\bea\label{eq:upnh}
u_p \approx \f{A_{\rm in}}{\sqrt{\D}} \,, \,\, A_{\rm in} = 2 \big|\mE_{\text{in}} - \Omega_H\mL_{\text{in}}\big| + \f{4r_+^p}{\eta_{\text{in}}}\left( \Omega_H-\Omega_F \right)^2 \,,
\eea
where $\Omega_H = a(2 r_+)^{-1}$ is the black hole angular velocity. 
The poloidal velocity therefore diverges as $\D^{-1/2}$ at the horizon, 
as expected because an infalling flow must become super-fast-magnetosonic near $r_+$ \citep{takahashi1990magnetohydrodynamic}. 
Then, the expressions for $u_t$ and $u_{\p}$ can be obtained by substituting Eq.~\eqref{eq:upnh} into the near-horizon expansion of Eq.~\eqref{eq:utup}:
\bea
\begin{aligned}
u_t \approx \f{-\mE_{\text{in}} \eta_{\text{in}} A_{\rm in} + ar_+^{p-2} \sqrt{k_S}\left(\Omega_F-\Omega_H\right)}{\eta_{\text{in}} A_{\rm in} + 4\left( \Omega_H-\Omega_F \right)^2 r_+^{p-1}} \,,\quad  u_\p \approx \f{\mL_{\text{in}} \eta_{\text{in}} A_{\rm in} - r_+^{p-1} \sqrt{k_S}\left(\Omega_F-\Omega_H\right)}{\eta_{\text{in}} A_{\rm in} + 4\left( \Omega_H-\Omega_F \right)^2 r_+^{p-1}} \,.
\end{aligned}
\eea
Hence, $u_t, u_\p$ asymptote to constants as $r \to r_+$, consistent with the setup in \citep{Hou:2024qqo}. The scalings of $u_p$ and $B_p$ imply that the inflow density approaches a constant at the horizon, $\rho_{\rm in} \to -\eta_{\rm in} r_+^{p-1}A_{\rm in}^{-1}$, consistent with the behavior shown in Fig.~\ref{fig:density}. 
Finally, the toroidal magnetic field component is obtained by substituting Eq.~\eqref{eq:upnh} into the near-horizon expansion of Eq.~\eqref{eq:Bphi02}:
\bea
B^{\p} \approx \f{1}{\D} \f{4\eta_{\text{in}} \left( \Omega_F-\Omega_H \right) \left( \mE_{\text{in}} - \Omega_F \mL_{\text{in}} \right)}{4r_+^{p-1}\left( \Omega_F-\Omega_H \right)^2 + \eta_{\text{in}} A_{\rm in} } \,.
\eea
This scaling indicates a highly toroidal configuration. Near the horizon, the winding numbers tend to $w_B,w_m \to 1$, as is illustrated in Fig.~\ref{fig:winding}. Although derived in the BL coordinates, this toroidal structure is not merely a coordinate artifact but reflects a genuine physical feature in rotating black holes. In particular, as the horizon is approached, it yields an increasingly radial EVPA pattern and produces a strongly spiraling image of a bright spot \citep{Chen:2024jkm,Hou:2024qqo}.
In addition, the regularity condition at the horizon imposes the following constraint on the flow variables:
\bea\label{eq:rpboundary}
\Omega_F^2 - \Omega_H \Omega_F + \f{1}{2}\eta_{\text{in}}\left( \mE_{\text{in}} +u_t \right)r_+^{1-p} = 0 \,,
\eea
which explicitly involves the fluid velocity; in the limit $\eta_{\rm in} \to 0$, Eq.~\eqref{eq:rpboundary} reduces to the well-known force-free Znajek condition.

\subsection{Asymptotic flow scalings}
\label{App:jetasym}

We now describe the asymptotic velocity and magnetic-field structure of the MHD outflow well outside the light cylinder, 
where gravity is negligible, and the dynamics is controlled by magnetic stresses and matter inertia. 
Throughout this section, we suppress the subscript ``out'' on all outflow quantities.
In the flat-spacetime limit, the temporal and azimuthal components of the four-velocity, obtained from the conserved quantities $\mE$ and $\mL$ defined in Eq.~\eqref{conservedquantities}, reduce to
\be
\label{eq:utupinf}
\bag
u_t = \mE \,\f{-1 + \Omega_F l + M_A^2}{1-R^2 \Omega_F^2- M_A^2} \,,  
\quad 
u_\p = \mE \,\f{R^2\Omega_F(1 - \Omega_F l)-M_A^2l}{ 1-R^2 \Omega_F^2 - M_A^2} \,.
\eag
\ee
The toroidal magnetic field is
\bea\label{eq:Bphiinf}
B^{\phi}= \eta \,\mE \,\f{-l + R^2 \Omega_F }{R^2 (1-R^2 \Omega_F^2 - M_A^2) } \,.
\eea
The corresponding wind equation is
\bea\label{eq:windinf}
\left(u_{p}^2+1\right)(k_0-M_A^2)^2 = \mE^2 \left( k_0k_2-2k_2M_A^2-k_4M_A^4\right)\,, \quad 
M_A^2  = \f{\eta u_p}{B_p} \,,
\eea
where $k_0 = 1- R^2\Omega_F^2$, $k_2= \left(1-\Omega_F l\right)^2$, and $k_4= -1 + l^2R^{-2}$. 
Under the condition in Eq.~\eqref{eq:infity1}, the wind equation depends on $\Omega_F, \g$, and $\mL$.
The parameter $\mL$ is then fixed by the regularity condition at the SS.
Specifically, requiring $B^{\phi}$ to remain regular at the SS, together with the corotation condition
$u^{\p}\big|_{\rm SS} = \Omega_F  u^t \big|_{\rm SS}$, gives
$\Omega_F \mL  = \mE -\sqrt{k_0}\big|_{\rm SS}$. 
Thus the wind equation may be parameterized by $\Omega_F$, $\g$, and $k_0\big|_{\rm SS}$ alone.
For brevity, we denote $k_0\big|_{\rm SS} = k_{\rm S}$. We have also neglected the subscript of the terminal Lorentz factor hereafter.

As $r \to \infty$, the flow becomes purely poloidal, with $u_p \to \sqrt{\g^2-1}$. 
We therefore introduce a dimensionless, large-distance variable $\tilde{R} = \Omega_FR \gg 1$, 
which measures the distance in units of the light-cylinder radius $R_l \sim \Omega_F^{-1}$.
To extract the approach to the terminal speed, we write $u_p = \sqrt{\g^2-1} - \d$ and solve Eq.~\eqref{eq:windinf} perturbatively in powers of $\tilde{R}^{-1}$.
This gives:
\bea\label{eq:expandup}
\d = \f{\sqrt{\g^4-(k_{\rm S}+2)\g^2 + 2\sqrt{k_{\rm S}}\,\g}}
{\sqrt{3}}\,\f{1}{\tilde{R}} - 
\f{\g^4 -3\g^2 + 2\sqrt{k_{\rm S}}\g - k_{\rm S} +1}{3\sqrt{\g^2 -1}} \,\f{1}{\tilde{R}^2}
\,+\, \mO (\tilde{R}^{-3}) \,.
\eea
Thus, to leading order, the poloidal velocity approaches its asymptotic value as $\big|u_p - \sqrt{\g^2-1}\,\big| \sim \tilde{R}^{-1}$.
The coefficient of this approach depends on the launching parameter $k_{\rm S}$, 
while the variable $\tilde{R}$ encodes the field-line rotation.
The temporal and azimuthal components follow from expanding Eq.~\eqref{eq:utupinf} and using $M_A^2 = \eta u_p /B_p$, yielding
\bea\label{eq:expandutup}
\begin{aligned}
 u_t = &-\g+\f{\sqrt{\g^2-1}\sqrt{\g^3-(k_{\rm S}+2)\g + 2\sqrt{k_{\rm S}}}}{\sqrt{3\g}}\,\f{1}{\tilde{R}} 
- \f{2\g^3 -3\sqrt{k_{\rm S}}\g^2 + (k_{\rm S} -1) \g + \sqrt{k_{\rm S}}}{3 \g^2}\, \f{1}{\tilde{R}^2}
\,+\, \mO (\tilde{R}^{-3}) \,, \nn \\
 u_{\p} = & \, - \f{\g^3 + \sqrt{k_{\rm S}}\g^2 - 2\sqrt{k_{\rm S}}}{\g^2\Omega_F}
+ \f{\sqrt{\g^2 -1}(\g^3-2\sqrt{k_{\rm S}})\sqrt{\g^3-(k_{\rm S}+2)\g + 2\sqrt{k_{\rm S}}}}{\sqrt{3}\g^{7/2}\Omega_F}\, \f{1}{\tilde{R}} \\
&\,-\f{2\g^6 + 3\sqrt{k_{\rm S}}\g^5 + (k_{\rm S}-1)\g^4 -9\sqrt{k_{\rm S}}\g^3 + (2\sqrt{k_{\rm S}}-2k_{\rm S}^{3/2})\g + 4k_{\rm S}}{3\g^5\Omega_F}\, \f{1}{\tilde{R}^2} \,+\, \mO (\tilde{R}^{-3}) \,.
\end{aligned}
\eea
At leading order, the angular momentum content is determined by $\g,\Omega_F$ and $k_{\rm S}$ through $\Omega_F\g^2 u_{\p} = - \left(\g^3 + \sqrt{k_{\rm S}}\g^2 - 2\sqrt{k_{\rm S}}\,\right)$. 
Based on the large-$\tilde{R}$ scalings of $u_p$ and $B_p$, the outflow density asymptotically approaches $\rho_{\rm out} \to 2\psi \,\eta_{\rm out}\tilde{R}^{-2}\left(\g^2-1\right)^{-1/2}$ at spatial infinity, consistent with Fig.~\ref{fig:density}. 
The toroidal magnetic field at large $\tilde{R}$ is obtained by expanding Eq.~\eqref{eq:Bphiinf}, yielding
\bea\label{eq:Bphiinf02}
B^{\p} \propto -\f{\g}{\Omega_F \tilde{R}^2\sqrt{\g^2 -1}} + \mO (\tilde{R}^{-3}) \,.
\eea
The corresponding toroidal field strength scales as $B_T = R |B^{\p}| \sim \tilde{R}^{-1}$, 
whereas the poloidal component scales as $B_p \sim \tilde{R}^{-2}$. 
This hierarchy implies that the field becomes increasingly toroidally dominated as the outflow propagates outward, with the field-line winding degree $w_B \to 1$. 
In contrast, since $u_p^2 \to \g^2 -1$ while $u_{\p}u^{\p} \sim R^{-2}$, 
the plasma motion remains predominantly poloidal, with $w_m \to 0$, as indicated in Fig.~\ref{fig:winding}. 

The force-free limit requires separate treatment, which is qualitatively different from MHD flows \citep{Shen:2026aye}.
In this limit, the jet is entirely magnetically dominated: the mass loading tends to zero, $\eta \to 0$, and the terminal Lorentz factor formally diverges, $\g \to \infty$. In this regime, the expansion $u_p = \sqrt{\g^2-1} - \d$ is no longer valid. This can be seen by expanding $\d$ in Eq.~\eqref{eq:expandup} in powers of $\g^{-1}$, which yields
\bea
u_p \approx \sqrt{\g^2-1} + \left(\g^2 - 1 - \f{k_{\rm S}}{2}\right) \f{1}{\sqrt{3}\tilde{R}} \,.
\eea
As $\g \to \infty$, the correction proportional to $\tilde{R}^{-1}$ grows faster than the finite-$\g$ leading term, signaling a breakdown of the expansion. 
Therefore, one must first take $\g \to \infty$ in the wind equation and only then perform the large-$\tilde{R}$ expansion.
In this limit, the wind equation reduces to a simple form:
\bea\label{eq:windFFE}
(1-\tilde{R}^2)\left(k_{\rm S}-u_p^2-1+\tilde{R}^2\right) = 0 \,,
\eea
which yields $u_p= \sqrt{k_{\rm S} - 1+\Omega_F^2x^2} \approx \tilde{R} +(k_{\rm S}-1)(2\tilde{R})^{-1}$. 
The leading term implies $u_p \sim \tilde{R}$ at large distances, i.e., unbounded acceleration, consistent with $\g \to \infty$. 
Taking the large-$\g$ limit of Eq.~\eqref{eq:utupinf}, the temporal and azimuthal components become
\bea\label{eq:utuphiFFE}
\begin{aligned}
u_t = &\,\f{\Omega_F \tilde{R} \sqrt{k_{\rm S}-1+\tilde{R}^2}-\sqrt{k_{\rm S}}}{1 - \tilde{R}^2}\approx  -\tilde{R} - \f{k_{\rm S}+1}{2\tilde{R}} \,, \\
u_\p = &\,  \f{\Omega_F^2\tilde{R} \sqrt{k_{\rm S}-1+\tilde{R}^2} + \sqrt{k_{\rm S}}\,\tilde{R}^2}{\Omega_F^2\left(1 - \tilde{R}^2\right)} \approx -\f{\tilde{R}}{\Omega_F} - \f{\sqrt{k_{\rm S}}}{\Omega_F} + \f{k_{\rm S}-1}{2\tilde{R}} \,.
\end{aligned}
\eea
At leading order, these satisfy $u_t = - \Omega_F u_\p$, while the imprint of jet launching ($k_{\rm S}$) appears in the subleading terms of $u_{\p}$. 
Finally, taking the same limit in Eq.~\eqref{eq:Bphiinf} yields $B^\p \propto \Omega_F^{-1} \tilde{R}^{-2}$, 
consistent with the large-$\g$ limit of Eq.~\eqref{eq:Bphiinf} and implying a purely toroidal magnetic field with $w_B \to 1$. 
Meanwhile, since  $u_p^2 \sim \tilde{R}^2$, $u_\p u^\p \sim 1$, the flow approaches a purely poloidal configuration at infinity, with $w_m \to 0$.

\medskip
\section{Synchrotron radiation and transport}
\label{App:SRT}

\subsection{Nonthermal emission profile}
\label{App:emission}

In weakly collisional plasma, processes such as turbulence, magnetic reconnection, or shocks accelerate part of the population, producing nonthermal electrons in the distribution. We model the local synchrotron emissivity produced by nonthermal electrons with a power-law energy distribution that works well in many cases \citep{EventHorizonTelescope:2021btj}, 
\begin{equation}
 F_p(\gamma) = \f{s-1}{\gamma_{\rm min}^{1-s}-\gamma_{\rm max}^{1-s}}\, \gamma^{-s}  \,,  \quad \text{for} \quad \gamma_{\rm min} \leq \gamma \leq \gamma_{\rm max} \,.
\end{equation}
Here $s$ is the spectral index, and $\gamma_{\rm min}, \gamma_{\rm max}$ specify the lower and upper cutoffs of the high-energy tail, determined by microscopic processes responsible for energy conversion in plasmas.  The lower cutoff is set by the peak of the Maxwell-J\"{u}ttner distribution: $\gamma_{\rm min} = 1 + f\left( \Theta_\text{e} \right) \Theta_\text{e}$, where $\Theta_\text{e} = k_BT_{\text{e}}/m_{\text{e}}$ is the dimensionless electron temperature. Since results are insensitive to $\gamma_{\rm max}$ in this range, the practical limit is often extended to infinity. 
The power-law index $s$ is prescribed using an empirical fit to local particle-in-cell (PIC) simulations of reconnection-driven heating \citep{Ball:2018icx}:
\begin{equation}\label{eq:PICp}
	p(\beta, \sigma_\text{M}) = 1.8 + 0.7 \sigma_\text{M}^{-0.5} + 3.7 \sigma_\text{M}^{-0.19} \tanh(23.4  \sigma_\text{M}^{0.26} \beta)   \,. 
\end{equation}
where $\sigma_M = b^2/\rho$ is the magnetization parameter. For the cold jet flows considered here, the plasma beta $\beta = p_{gas}/p_{B}$ is set to zero.
The collective synchrotron emissivity from the power-law eDF can be obtained as \citep{1979Lightman}
\bea\label{eq:eDFemission}
j_{\nu} \propto \, && \,\rho \, \nu^{(s-1)/2} \left( \tilde{b} \,|\sin{\a_B}| \right)^{(s + 1)/4} \,,
\eea
where $\nu = -k_{\m}u^{\m}$ is the emitted photon frequency in the co-moving fluid frame, with $k^{\m}$ the wave vector. The magnetic field strength, $\tilde{b} = \sqrt{b^\m b_\m}$, is evaluated in the co-moving fluid frame as well; 
$\a_B = \cos^{-1}( \n^{-1} \tilde{b}^{-1} b^{\m}k_{\m})$ is the pitch angle between $k^{\m}$ and $b^{\m}$. 
The overall coefficient in Eq.~\eqref{eq:eDFemission} depends on the incidence direction of photons into the emission layer and is computed by contracting the wave vector with the tangent to the stream function (the principal thickness direction), $\propto |k^\mu \partial_{\mu}\psi|$. For details, see \citep{Gelles:2024tpz}.

\begin{figure}[htbp]
\centering
      \includegraphics[width=2.6in]{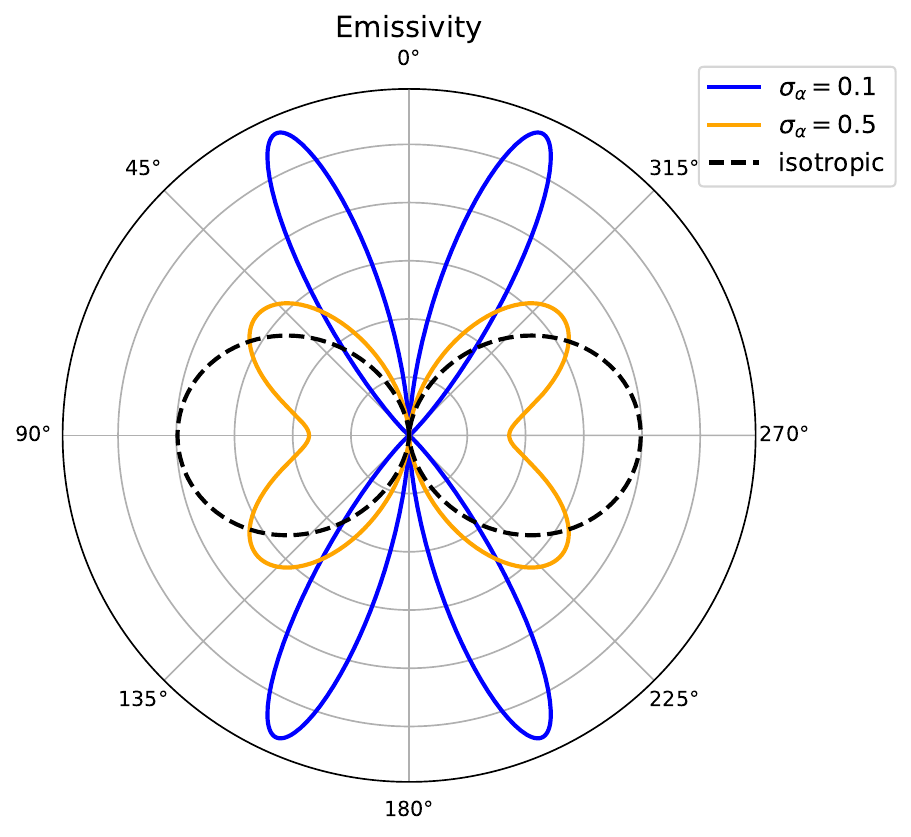}
	\caption{Schematic illustration of the pitch-angle dependence of the synchrotron emissivity for isotropic and anisotropic eDFs, shown in the comoving fluid frame.}
	\label{fig:anisoeDF}
\end{figure}

Embedded in dynamically strong magnetic fields, the electrons easily develops anisotropies. 
Although the eDF is generally gyrotropic, it can exhibit pronounced anisotropy between directions parallel and perpendicular to the field \citep{kulsrud1983mhd}. 
A simple example is the normalized Gaussian function \citep{Lai:2025yeq, Zhou:2025moa},   
\bea\label{beam1}
&& G_{b}(\alpha) = \frac{1}{X} \exp\left(-\frac{(\cos\alpha - \cos\alpha_0)^2}{2\sigma^2}\right)\,, \\
&& X = \sqrt{2\pi^3 \sigma^2}\left[\text{erf}(t_2) - \text{erf}(t_1)\right] \,, \nn \\
&& t_2 = \frac{1}{\sqrt{2\sigma^2}}(1 - \cos\alpha_0) \,, \quad t_1 = -\frac{1}{\sqrt{2\sigma^2}}(1 + \cos\alpha_0) \,, \nn
\eea
where $\alpha_0$ and $\sigma$ control the center and width of the Gaussian function. 
PIC simulations have shown that accelerated electrons during reconnection events can develop a beam-like eDF along the local magnetic field \citep{comisso2022ion, Comisso:2023ygd}. Thus, Eq.~\eqref{beam1} with $\alpha_0 = 0$ can describe how particles are preferentially accelerated along field lines. 
We multiply the emissivity by such a beam-like function to model the potential pitch-angle anisotropy, yielding
\bea\label{eq:eDFemission02}
j_{\nu} \propto \, && \,\rho \, \nu^{(s-1)/2} e^{-\f{1}{2\sigma^2}(\cos\a_B - 1)^2} \left( \tilde{b} \,|\sin{\a_B}| \right)^{(s + 1)/4} \\ \nn 
&& + \,\left(\a_B \to \a_B + \pi\right) \,.
\eea
We illustrate this pitch-angle dependence in Fig.~\ref{fig:anisoeDF}, which clearly demonstrates the effect of the anisotropy function in Eq.~\eqref{beam1} on the emissivity. Specifically, it shifts the peak of $j_\nu$ away from $\alpha_B=\pi/2$ toward the two poles. For $\sigma=0.1$, the peaks are located at approximately $20^\circ$ and $160^\circ$, already very close to the polar directions.

We focus on the optically thin regime, which is true for $\gtrsim 86$ GHz jet observations \citep{Hada:2013yla, Hada:2015okc, lee2016interferometric}, the emitted linear polarization vector is perpendicular to the magnetic field \citep{1979Lightman}
. 
It can be expressed covariantly as \citep{Chen:2025ysv}
\bea\label{eq:polvecdef}
f^{\mu} \propto \epsilon^{\mu\nu\rho\sigma} u_{\nu} B_{\rho} k_{\sigma} \,.
\eea
Under ideal MHD, we can rewrite the polarization vector covariantly in terms of the drift velocity \citep{McKinney:2006tf}. The result is
\bea\label{eq:fmure}
f^{\m} \propto  \,\epsilon^{\m\n\a\b} \eta_{\n} k_{\a} \mathcal{B}_{\b} + z_{\perp} \left( \mathcal{E}^{\m} + \eta^{\m} \mathcal{E}^{\rho}k_{\rho} \right) \,, 
\eea
where $\epsilon^{\m\n\a\b}$ is the Levi-Civita tensor, $\eta^{\m}$ is the spacetime normal vector; $ z_{\perp} = v_{\perp}/v^{\text{drift}}_{\perp}$ quantifies the flow velocity relative to the drift motion perpendicular to $\mathcal{B}^{\m}$ \citep{Chael:2023pwp, Gelles:2026mxg}. 
Far from the black hole, we have $\vec{f} \sim \vec{k}\times \vec{B} + z_{\perp} \vec{E}$, with $\vec{E} \approx r \sin{\t} \,\Omega_F \,\hat{e}^{(\p)}\times \vec{B}$ that reflects the ``aberration'' induced by high speed motion.

In the optically thick regime, like $\lesssim 43$ GHz observations for M87* \citep{macdonald1968observations, junor1999formation, Asada:2011dr}, preferential absorption of the perpendicular polarization mode causes the escaping radiation to have a polarization vector $f^{\m}$ parallel to the magnetic field \citep{1979Lightman, Tsunetoe:2024uzh}. 

\subsection{Discrete GRRT}
\label{App:DGRRT}

Because the local emission is highly sensitive to the pitch angle $\a_B$ (within the acceleration region in jet), even moderate gravitational lensing can produce appreciable effects, making a GRRT treatment necessary. This is particularly evident in the counter-jet-dominated emission region.

It is well known that, in Kerr spacetime, the structure of null geodesics is highly tractable. In Boyer-Lindquist coordinates $\left(t,r,\t,\p\right)$, the null geodesic equations reduce to first-order form in the $r$ and $\theta$ directions:
\bea
\begin{aligned}\label{eq:Kerrnulleq}
&k_{\mu} = \left(  -1, \eta_r \frac{\sqrt{\mathcal{R}(r)}}{\Delta}, \eta_\theta \sqrt{\Theta(\theta)}, l \right) \,, \\
&\mathcal{R} = \left(r^2 + a^2 -a l\right)^2 - \Delta \left[ \eta +  \left( l - a \right)^2 \right]  \,,\\
&\Theta =  \eta + a^2\cos^2{\theta} -l^2\cot^2{\theta} \,,
\end{aligned}
\eea
where $\Delta = r^2-2Mr+a^2$, and $\eta_r$ and $\eta_\theta$ denote the signs of $k_r$ and $k_\theta$, respectively. The quantities $l$ and $\eta$ are the impact parameters associated with the conserved angular momentum and Carter constant. Note that we have set the photon energy to unity by rescaling the affine parameter. Using Eq.~\eqref{eq:Kerrnulleq}, the null geodesic equations can be recast into the following lens equations \citep{Vazquez:2003zm}:
\bea
\begin{aligned}
& \fint^{r_o}_{r_s} \frac{\md r}{\eta_r\sqrt{\mathcal{R}}} = \fint^{\theta_o}_{\theta_s} \frac{\md \theta}{\eta_\theta\sqrt{\Theta}} \,,  \\
& \phi_o - \phi_s = a \fint^{r_o}_{r_s} \frac{2M r - a l}{\eta_r\Delta\sqrt{\mathcal{R}}} \md r + l \fint^{\theta_o}_{\theta_s} \frac{\csc^2{\theta}}{\eta_\theta\sqrt{\Theta}} \md\theta \,, \\
& t_o - t_s = \fint^{r_o}_{r_s} \frac{r^2\Delta + 2Mr(r^2 + a^2 -a l)}{\eta_r\Delta\sqrt{\mathcal{R}}} \md r  + a^2 \fint^{\theta_o}_{\theta_s} \frac{\cos^2{\theta}}{\eta_\theta\sqrt{\Theta}} \md\theta,
\end{aligned}
\eea
where $\left( t_s,r_s,\theta_s,\phi_s \right)$ and $\left( t_o,r_o,\theta_o,\phi_o \right)$ denote the source and observer positions, respectively. This formalism allows us to determine the mapping accurately from a source position to its image on the observer's screen.

To produce synthetic images, we must solve the radiative transfer equations along null geodesics. The local Stokes parameters are defined as $\Mc{S} \in \{I,Q,U,V\}$, where $I$ is the total intensity, $Q$ and $U$ denote the linear polarization, and $V$ denotes the circular polarization. Note that these Lorentz-invariant quantities are not the directly observed frequency-dependent Stokes parameters; rather, they are related to them through $X = \nu^{-3}X_\nu$. The transfer equation then takes the form \citep{Broderick:2003fc, Shcherbakov:2010kh}
\bea\label{eq:RTe01}
\dfrac{\md}{\md \lambda} \Mc{S} =  \dfrac{1}{\nu^2}j_{\n}^{(\Mc{S})} - \nu \Mc{R}_\nu \Mc{S}\,,
\eea
where $j_{\n}^{(\Mc{S})}$ denotes the polarized emissivities, $\Mc{R}_\nu$ is a matrix that contains the absorption and Faraday rotation coefficients for the Stokes parameters, and $\lambda$ is the affine parameter along the null geodesic.

The polarized image is obtained by solving Eq.~\eqref{eq:RTe01} along all photon trajectories connecting the emission source to the observer. For a geometrically and optically thin emission layer, the observed intensity $I^{(o)}_\nu$ and EVPA can be further evaluated as a sum over multiple crossings of the layer:
\bea\label{eq:RT01}
I^{(o)}_\nu = \sum_i g_i^3 I_{\n,i} \,,\,\,\, \text{EVPA} = \f{1}{2} \arctan{\left(\f{\sum_i U_{\n,i}}{\sum_j Q_{\n,j}}\right)}  \,,
\eea
where $i$ labels the $i$-th crossing of the emission layer, and $g_i$ is the corresponding redshift factor. Although the EVPA depends on the choice of directional convention, it uniquely specifies the polarization pattern measured on the observer's screen \citep{EventHorizonTelescope:2021bee,EventHorizonTelescope:2021srq}.

The subsequent evolution of the polarization vector generated at each layer crossing along the null geodesic is governed by parallel transport. Since Kerr spacetime is of Petrov type D, the parallel-transport equation can be simplified by introducing a conserved quantity. According to the Walker-Penrose (WP) theorem \citep{Walker:1970un}, a real normalized parallel-transported vector $f^\mu$ defines a complex scalar
\bea\label{eq:WPcon}
\begin{aligned}
&\kappa = \left( r - ia\cos{\theta} \right) \left[ A -  i B \right] \,, \\
& A = 2 k^{[t}f^{r]} + 2a \sin^2{\theta}\, k^{[r}f^{\phi]} \,, \\
& B =  2\sin{\theta} \left[ \left(r^2+a^2\right)k^{[\phi}f^{\theta]} - a k^{[t}f^{\theta]} \right]  \,,
\end{aligned}
\eea
which is conserved along the geodesic, $k^\mu\partial_{\mu}\kappa = 0$. Therefore, the polarization phase associated with the $i$-th crossing in Eq.~\eqref{eq:RT01} is encoded in the WP constant as
\bea\label{eq:UiOi}
\f{1}{2} \arctan{\left(\f{U_{\n,i}}{Q_{\n,i}}\right)} = \arctan\left[\f{\m \,\text{Re}(\kappa_i) - y \,\text{Im}(\kappa_i)}{y \,\text{Re}(\kappa_i)  + \m \,\text{Im}(\kappa_i)}\right]\,, 
\eea
where $\m = -\left( x + a\sin{\t_o} \right)$, and $\{x,y\}$ are the Bardeen coordinates on the observer's screen, determined by the photon's impact parameters \citep{cunningham1973optical}; $\kappa_i$ is the WP constant constructed via Eq.~\eqref{eq:WPcon} at the $i$-th crossing. We see that Eq.~\eqref{eq:UiOi} is frequency-independent, because the linear polarization vector is determined geometrically by the local magnetic field and photon wave vector.

\subsection{SRRT}
\label{App:SRRT}

In flat spacetime, light travels along straight lines, so the mapping between a point on the image plane and its corresponding emission point on a jet layer can be obtained analytically. To make this construction explicit, we adopt Cartesian coordinates $(x',y',z)$ centered on the black hole and place the observer in the $y'$--$z$ plane (the prime notation is used to distinguish these from the screen coordinates $x,y$). The line of sight is then
\[
\hat{o} = \left(0,\sin\theta_o,\cos\theta_o\right),
\]
where $\theta_o \in \left[-\pi,\pi\right] $ is the viewing angle. The observer's image plane is parameterized by the Cartesian coordinates $(x,y)$, with basis vectors
\[
\hat{e}_{(y)} = -\hat{e}_{(x')} = (-1,0,0), 
\qquad
\hat{e}_{(x)} = \hat{e}_{(y')} \times \hat{o}
= \left(0,\cos\theta_o,-\sin\theta_o\right).
\]
As the image plane is well defined by parallel light rays for the distance observer, we neglect the observer's specific location and assume free parallel transport of the vectors defined on the image plane.
An image point $\mathcal{P}$ can then be related to
\begin{equation}
\vec{\mathcal{P}}
= b\left(-\sin\varphi,\cos\varphi\cos\theta_o,\cos\varphi\sin\theta_o\right),
\end{equation}
where $b=\sqrt{x^2+y^2}\in(0,\infty)$ is the image-plane radius and $\varphi \in [0,2\pi)$ is the polar angle measured from the $x'$-axis. For fixed $b$, the corresponding light rays form an inclined cylinder of radius $b$ whose axis is aligned with $\hat{o}$.
The straight line passing through the image point labeled by $(b,\varphi)$ is parameterized by
\begin{equation}
x'=-b\sin\varphi,\qquad
y'=h\sin\theta_o+b\cos\varphi\cos\theta_o,\qquad
z=h\cos\theta_o+b\cos\varphi\sin\theta_o,
\end{equation}
where $h\in(-\infty,\infty)$ denotes the distance along the line of sight. We then determine its intersections with the jet layer, which is described by
\begin{equation}
\psi=r^p\left(1-|\cos\theta|\right)
\,\approx\, \frac{x'^2+y'^2}{2|z|^{\,2-p}}
\end{equation}
in the large-distance region where the SRRT approximation applies. This yields
\begin{equation}
2\psi\left|h\cos\theta_o+b\cos\varphi\sin\theta_o\right|^{2-p}
=
b^2\sin^2\varphi+
\left(h\sin\theta_o+b\cos\varphi\cos\theta_o\right)^2,
\end{equation}
from which $h$ can be solved. 
Depending on the geometry, this equation typically admits zero or two real roots. The two-root case is generic; it corresponds to intersections with both the forward and counter jets for a nearly face-on view, or to two intersections with the same cone for a nearly edge-on view. 
If the ray is tangent to the jet surface, the equation admits a single root. 
For an inclined observer and a small impact parameter, three or four roots may also occur, with two roots on one cone and one or two on the other.
In this way, the image point $(b,\varphi)$ is mapped to the emission point $\vec{s}$,
\begin{equation}
\vec{s}
=\vec{\mathcal{P}}+h\,\hat{o}
=
\left(
-b\sin\varphi,\,
b\cos\varphi\cos\theta_o+h\sin\theta_o,\,
b\cos\varphi\sin\theta_o+h\cos\theta_o
\right).
\end{equation}

When producing Fig.~\ref{fig:windingeta2} in the main text, we verified that the SRRT performs very well in the asymptotic region, i.e., for $b > 200\,M$. In this region, only the forward-jet emission is included, since the counter-jet emission is strongly suppressed by Doppler deboosting.

\medskip
\section{Asymptotic EVPA viewed face-on}
\label{App:infevpa}

We next examine how the velocity and magnetic-field structures in the far zone ($X \gg 1$) determine the asymptotic polarization pattern.
The polarization vector at the emission point is expressed covariantly as $f^{\mu} \propto \epsilon^{\mu\nu\rho\sigma} u_{\nu} B_{\rho} k_{\sigma}$. In the far zone, where general relativistic effects can be neglected, light propagates along straight lines. For an on-axis observer located at the north pole, the wave vector reduces to
\bea\label{eq:wavevector}
k_{\m} = \left( -1, \sqrt{1-b^2r^{-2}}, -b, 0 \right)\,,
\eea
where the impact parameter $b$ denotes the horizontal distance from the emission point to the jet axis; for an on-axis view, $b = x$.  
For a jet layer described by the stream function $\psi={\rm const} = r^p\left( 1-|\cos{\t}| \right)$, 
inverting this relation yields the connection between the cylindrical radius $x$ and the spherical radius $r$ along the jet, 
$b = R = \sqrt{\psi r^{2-2p} \left( 2r^p - \psi \right)}$. 
In the asymptotic regime, this geometry implies $\sin\theta \simeq \sqrt{2\psi}\, r^{-p/2}$ and  $R\simeq \sqrt{2\psi}\, r^{1-p/2}$.

For a finite terminal Lorentz factor, the asymptotic flow velocity is given by Eqs.~\eqref{eq:expandup} and \eqref{eq:expandutup}. The poloidal magnetic field follows directly from the stream function and the frozen-in condition, 
$B^r u^\t = B^\t u^r$, yielding $B^r = r^{p-2} {\rm sign} \left( \cos{\t} \right)$ and $B^\t = -p\psi /\left(r^3\sin{\t}\right)$. 
Using Eq.~\eqref{eq:Bphiinf02}, the toroidal component becomes
\bea
B^\phi=
-\frac{\gamma \Omega_F}{\sqrt{\gamma^2-1}}\, r^{p-2}
+\mO\!\left(r^{3p/2-3}\right) \,.
\eea
Substituting $k^{\m}$, $u^{\m}$, and $B^{\m}$ into Eq.~\eqref{eq:polvecdef} and expanding $f^\mu$ in terms of $r^{-1}$, we can obtain the asymptotic expression for the polarization vector, parameterized by $\{\psi,\Omega_F,k_{\rm S},p,\g\}$.  
Only the transverse components $f^\theta$ and $f^\phi$ determines the observed electric-vector position angle (EVPA).
To leading order, we have
\bea\label{eq:fthetafphimhd}
\begin{aligned}
f^\t
&=
2\gamma
\left(
\frac{\gamma}{\sqrt{\gamma^2-1}}-1
\right)
\psi \Omega_F
+\mO\!\left(r^{p/2-1}\right) ,\\
f^\phi
&=
\left(1-\frac{p}{2}\right)\gamma \sqrt{2\psi}\, r^{p/2-1}
+\mO\!\left(r^{p-2}\right) .
\end{aligned}
\eea
Here, the overall multiplicative factor common to both components has been omitted, as it does not affect the EVPA.
Eq.~\eqref{eq:fthetafphimhd} shows that $f^\theta$ approaches a constant, whereas $f^\phi$ is asymptotically suppressed. 
The EVPA is defined as $\mathrm{EVPA}= \arctan\left(-\sin\t\,f^\phi/f^\t\right)$ \citep{Chen:2024jkm}, where the variable $\sin\theta \sim r^{-p/2}$ exactly compensates the $r^{p/2-1}$ scaling of $f^\phi$. As a result, $\sin\theta\,f^\phi/f^\theta \sim r^{-1}$ at leading order, independent of the collimation index $p$. 
Expanding to higher orders in $r^{-1}$, we obtain
\bea\label{eq:evpamhdnlo}
\mathrm{EVPA}\big|_{\rm MHD}
&\approx \,&
\frac{(2-p)\sqrt{\gamma^2-1}}{2\Omega_F (\sqrt{\gamma^2-1}-\gamma)} \frac{1}{r}
-\frac{\psi \left [4\gamma + (4 - 4p -2 p^2 + p^3)\sqrt{\gamma^2-1} \right ] \sqrt{\gamma^2-1} }{8\Omega_F(\sqrt{\gamma^2-1}-\gamma)^2}
\frac{1}{r^{p+1}} \nn\\
&&+ \frac{(2-p)\sqrt{2\sqrt{k_{\rm S}} -( 2+k_{\rm S}) \gamma  + \gamma^3} \sqrt{\gamma^2-1}}
{2\sqrt{6}\, \gamma^{3/2} \, \psi^{1/2}\, \Omega_F^2 (\sqrt{\gamma^2-1}-\gamma)^2}
\frac{1}{r^{2-p/2}}
\,.
\eea
Therefore, for a general MHD outflow viewed face-on, the EVPA approaches zero at large radius. This leading-order result is independent of the collimation index $p$. We note, however, that for certain exceptional values of $p$, the full coefficient of $f^\theta$ vanishes. In such cases, the ratio
$\sin\theta\,f^\phi/f^\theta$ becomes singular, and the expansion in Eq.~\eqref{eq:evpamhdnlo} breaks down. These special cases require a separate asymptotic treatment and will be addressed in our future work.

The force-free polarization limit must again be treated separately. As shown in Sec.\ref{App:jetasym}, the force-free asymptotics cannot be recovered by taking the limit $\gamma\to\infty$ of the finite-$\gamma$ expressions. 
Combining the force-free velocity solution in Eqs.\eqref{eq:windFFE} and \eqref{eq:utuphiFFE} with the magnetic-field scalings, we obtain
\bea\label{eq:fthetafphiffe}
\begin{aligned}
f^\theta=
\sqrt{2\psi}\, r^{p/2-1}
+\mO\!\left(r^{p-2}\right) \,, \quad 
f^\phi=
(2-p)\psi\Omega_F
+\mO\!\left(r^{p/2-1}\right) \,,
\end{aligned}
\eea
consistent with the result in \citep{Gelles:2024tpz}.
This represents the reverse hierarchy of the finite-$\gamma$ case: in the force-free branch, $f^\phi$ asymptotes to a constant, whereas $f^\theta$ decays.
Consequently, the large-radius EVPA is governed by
\bea
\sin\t\,\frac{f^\phi}{f^\theta}
\sim
(2-p)\psi\Omega_F\, r^{1-p}\,.
\eea
The sign of $(1-p)$ then determines the asymptotic polarization direction.
For $p<1$, the above ratio diverges, and the EVPA approaches $-\pi/2$:
\bea\label{eq:evpaffe_p_lt_1}
\mathrm{EVPA}\big|_{\rm FFE} \to
-\frac{\pi}{2}
+
\frac{1}{(2-p)\psi\Omega_F}\, r^{p-1}
+\mO\!\left(r^{3p-3}\right) .
\eea
For $p=1$, the ratio tends to a constant, hence $\mathrm{EVPA}\big|_{\rm FFE} \to
-\arctan(\psi\Omega_F)
+\mO(r^{-1/2})$.
For $p>1$, the ratio vanishes, so that $\mathrm{EVPA}\big|_{\rm FFE} \to
-(2-p)\psi\Omega_F\, r^{1-p}
+\mO\!\left(r^{3-3p}\right)$.
Therefore, the EVPA provides a clean diagnosis of the order in which the limits are taken. Finite-magnetization MHD flows always approach a vanishing face-on EVPA, whereas the force-free branch retains a geometry-dependent asymptotic polarization angle. This non-commutativity between the finite-$\g$ and force-free limits constitutes the central result of our analysis.

\medskip
\section{Additional polarization patterns}
\label{App:suppimage}

\subsection{Inclination effects}
\label{App:inclinaiton}

\begin{figure}[htbp]
\centering
\includegraphics[width=6in]{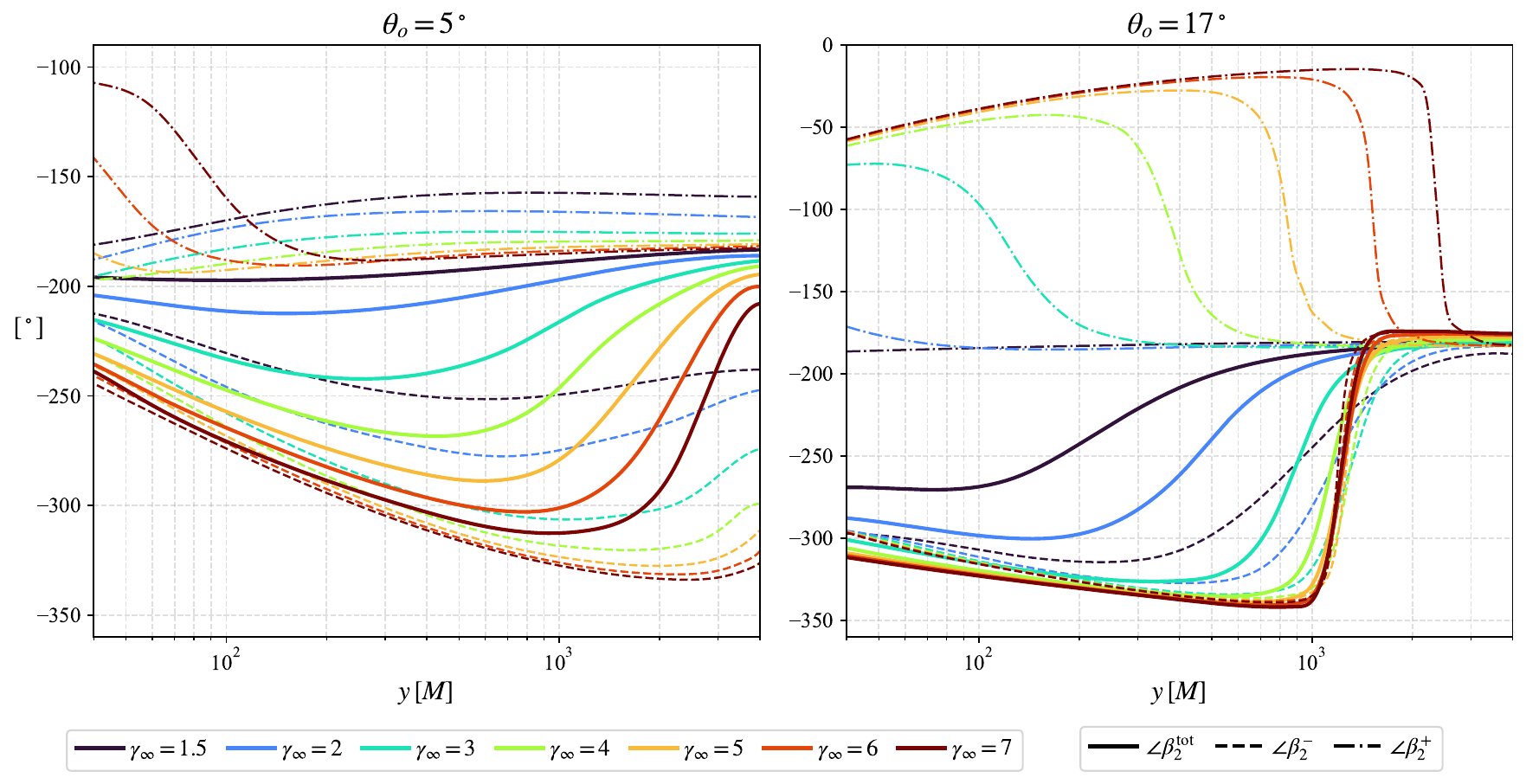}
	\caption{Mean polarization angles averaged over different segments along the $x$-axis in the forward jet ($y>0$), viewed at $\t_o = 5^{\circ}$ and $\t_o = 17^{\circ}$ for different values of $\g_\infty$. The black hole spin is fixed at $a = 0.5$, and the jet collimation index is $p = 1$. The solid, dashed, and dash-dotted curves show the variations of $\arg(\b^{\rm tot}_2)$, $\arg(\b^-_2)$, and $\arg(\b^+_2)$, respectively.}
	\label{fig:inclibeta}
\end{figure}

In this section, we present the polarization patterns as seen by an inclined observer. Because the jet is highly collimated, even a modest viewing angle can break the axial symmetry of the EVPA at locations far from the image center, making the usual azimuthal Fourier decomposition no longer well suited to this geometry. We therefore redefine a ``mean EVPA'' by averaging over a segment along the $x$-axis:
\bea
\beta_2(y) = \f{\int^{x_{\rm max}}_{x_{\rm min}} \left[Q_o(x,y) + i U_o(x,y) \right] e^{-2i\varphi} \, \md x }{\int^{x_{\rm max}}_{x_{\rm min}} I_o(x,y) \, \md x }\,,
\eea
where $x_{\rm min}$ and $x_{\rm max}$ denote the left and right integration boundaries. In particular, if these boundaries are chosen to coincide with the two jet limbs, i.e., the edge contours on both sides, this definition reduces to a quantity closely analogous to that used in the nearly face-on case, since $\md x \simeq y\, \md \varphi$ for highly collimated jet images on the screen. The solid curves in Fig.~\ref{fig:inclibeta} show the SRRT results of $\arg(\beta^{\rm tot}_2)$ for different $\g_\infty$, viewed at $\t_o = 5^{\circ}$ and $\t_o = 17^{\circ}$.

We find that, as in the face-on case, a pronounced EVPA bump persists for flows with high Lorentz factors. Its physical origin, however, is different in the inclined geometry. In this case, the bump arises from the combined emission from the foreground and background sides of the same jet layer along the line of sight. The background side can naturally have a larger pitch angle than the foreground side (without the aid of lensing), because the corresponding emission originates closer to the black hole, where the field lines have a smaller local slope. At sufficiently large distances from the black hole, however, Doppler beaming from the foreground side becomes dominant. The observed bump therefore reflects the competition between the pitch-angle effect and Doppler boosting. By contrast, in the face-on geometry, the bump is instead produced by the analogous competition between the forward and counter jets in the double-cone structure.

As the plasma mass loading increases, this bump feature is suppressed, just as in the face-on case. The underlying reason is again the same: heavier plasma enhances the MHD effect, leading to stronger field-line winding and a reduced pitch-angle contrast between the foreground and background sides. We therefore conclude that the imprint of plasma mass loading on the EVPA profile remains robust against moderate inclination.

As $y \to \infty$, $\arg(\beta^{\rm tot}_2)$ converges to $-180^{\circ}$, corresponding to a horizontal EVPA pattern parallel to the $x$-axis. This differs from the face-on case, where $\arg(\beta^{\rm tot}_2) \to 0^{\circ}$. The difference is purely geometric: for a distant inclined observer located outside the jet funnel (bounded by the sheath), the averaged EVPA is dominated by the bright jet limb, which typically accumulates more intensity than the central region because of its longer path length through the emitting flow \citep{Papoutsis:2022kzp}. As can be seen from the polarized images in Fig.~\ref{fig:jetimage} (and Fig.~\ref{fig:anisoimage} in Appendix~\ref{App:aniso}), the jet limb on the screen tends to exhibit EVPAs nearly parallel to the edge contour, whereas the central region shows a more vertical EVPA pattern. Consequently, the mean EVPA integrated along the $+x$-axis is largely controlled by the limb emission.

To further characterize the EVPA asymmetry between the two sides with respect to the $y$-axis, we define $\beta_2^{+}(y)$ and $\beta_2^{-}(y)$ as the corresponding averages taken over the positive and negative $x$-sides, respectively. These quantities are advantageous in that they capture the dominant polarization properties on both sides, while also highlighting the antisymmetry about the $y$-axis induced by relativistic aberration of the emitting plasma. The results are summarized in Fig.~\ref{fig:inclibeta}.

As $\g_\infty$ increases, the profile of $\beta_2^{+}(y)$ evolves from a smooth variation to a steeper one. For $\t_o = 17^{\circ}$, a pronounced bump appears and shifts toward $0^{\circ}$, indicating the emergence of a radial EVPA pattern. By contrast, $\beta_2^{-}(y)$ exhibits behavior markedly different from that of $\beta_2^{+}(y)$: for $\t_o = 5^{\circ}$, it generally decreases from $\sim -200^{\circ}$ to smaller values, before returning to $-180^{\circ}$ at very large distances. For $\t_o = 17^{\circ}$, $\beta_2^{-}(y)$ becomes more similar in shape to $\beta_2^{+}(y)$, but with the opposite sign; its bump is also steeper than that of $\beta_2^{+}(y)$. Overall, the asymmetry between these two quantities implies a generally nonzero net polarization when averaged over the entire horizontal axis.

\begin{figure}[htbp!]
\centering
\includegraphics[width=6.6in]{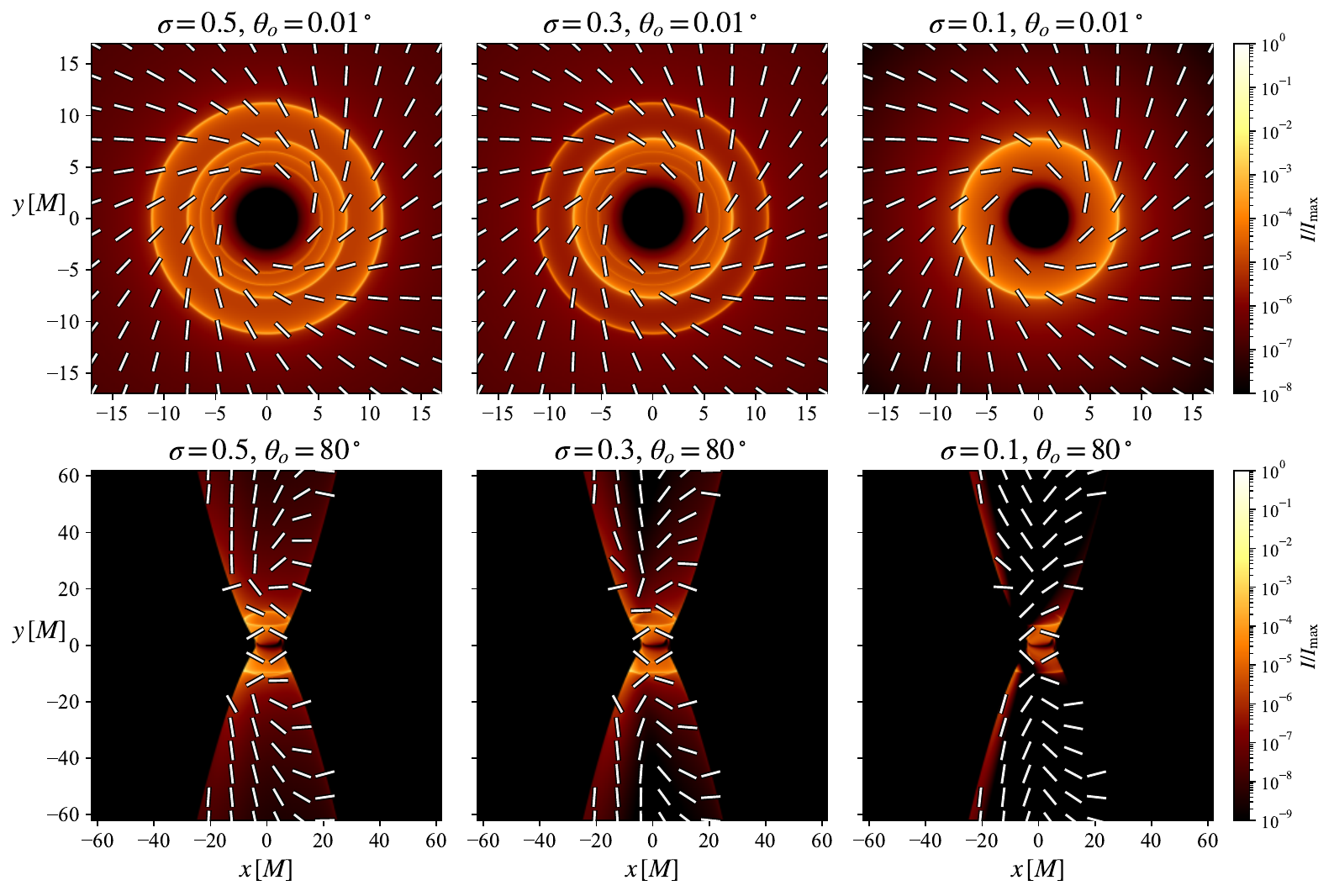}
	\caption{Intensity maps overlaid with EVPAs for anisotropic eDF emission, computed with $a = 0.5$, $p = 0.75$, and $\g_{\infty} = 1.5$. The left, middle, and right panels correspond to bi-beam-like eDFs with $\sigma = 0.5$, $0.3$, and $0.1$, respectively. The top and bottom rows show the results for a face-on observer ($\t_o = 0.01^{\circ}$) and a nearly edge-on observer ($\t_o = 80^{\circ}$), respectively.}
	\label{fig:anisoimage}
\end{figure}

\subsection{Anisotropic eDF effects}
\label{App:aniso}

In the main text, we assume an isotropic electron distribution in the fluid rest frame. This is a reasonable approximation for the disk, but it need not hold in the jet. In dynamically strong magnetic fields, weakly collisional electrons can readily develop velocity-space anisotropy. Although the eDF remains gyrotropic, its parallel and perpendicular components can differ substantially \citep{kulsrud1983mhd}. Previous studies have shown that such anisotropy can modify the synchrotron spectrum \citep{Lai:2025yeq} and may leave observable imprints in millimeter-band images \citep{Zhou:2025moa, Tsunetoe:2025crz, Glaser:2026sgs}. To model pitch-angle anisotropy, we multiply the power-law emissivity by a beam-like Gaussian function (Eq.~\eqref{eq:eDFemission02}), where $\sigma$ controls the angular width.

For an optically thin jet layer, the local polarization vector is perpendicular to both the magnetic field and the wave vector, and is therefore insensitive to the eDF anisotropy. However, a given light ray may intersect the jet layer multiple times. The observed polarization is then obtained from the total Stokes parameters, weighted by the intensity at each crossing. Consequently, eDF anisotropy affects the polarization only in cases with multiple jet crossings.

For a nearly edge-on view, rays outside the photon ring typically intersect the jet sheath twice, while only those grazing the sheath cross it once. As a result, in the polarized image, the central region (the jet spine, observationally) is more strongly affected by eDF anisotropy, whereas the jet edge (the sheath) remains largely unchanged. For a nearly face-on view, rays outside the photon ring intersect the forward jet once, and the counter-jet once, and such a characteristic double-cone structure can encode the eDF anisotropy. However, Doppler boosting suppresses emission from the counter-jet and enhances that from the forward jet; at sufficiently large scales ($x \gtrsim 100 M$) the image is therefore dominated by the forward jet.

Fig.~\ref{fig:anisoimage} shows polarized images produced with different levels of anisotropy. Comparing the EVPA distribution for  $\sigma = 0.3$ with the isotropic case, we find that the EVPA remains nearly unchanged across the image, despite the double-crossing effect noted above. For a more strongly anisotropic case ($\sigma = 0.1$), differences emerge: in the face-on view, the EVPA deviates within  $b \lesssim 10 M$, while in the edge-on view, the central region exhibits a qualitatively different EVPA structure compared to the isotropic case.       
We therefore conclude that for moderate anisotropy ($\sigma \geq 0.3$), the eDF anisotropy can be neglected when evaluating the global EVPA pattern. Only in strongly anisotropic cases (e.g., $\sigma = 0.1$) does it produce noticeable changes, primarily in the jet spine for nearly edge-on observers.

\bibliography{MHDJet}

@article{Glaser:2026sgs,
    author = "Glaser, Felix and Fromm, Christian M. and Ricci, Luca and Mizuno, Yosuke and Kadler, Matthias and Mannheim, Karl and Janssen, Michael",
    title = "{Probing anisotropic particle acceleration and limb-brightening in Centaurus A's jet}",
    eprint = "2603.26239",
    archivePrefix = "arXiv",
    primaryClass = "astro-ph.HE",
    month = "3",
    year = "2026"
}

@article{Chael:2026fhf,
    author = "Chael, Andrew and Lupsasca, Alexandru and Wong, George N. and Gelles, Zachary and Quataert, Eliot",
    title = "{Black Hole Polarimetry III: Universal Polarization of Synchrotron Radiation at the Horizon}",
    eprint = "2606.12518",
    archivePrefix = "arXiv",
    primaryClass = "astro-ph.HE",
    month = "6",
    year = "2026"
}

@article{Tsunetoe:2022ktx,
    author = "Tsunetoe, Yuh and Mineshige, Shin and Kawashima, Tomohisa and Ohsuga, Ken and Akiyama, Kazunori and Takahashi, Hiroyuki R.",
    title = "{Investigating the Disk{\textendash}Jet Structure in M87 through Flux Separation in the Linear and Circular Polarization Images}",
    eprint = "2202.12904",
    archivePrefix = "arXiv",
    primaryClass = "astro-ph.HE",
    doi = "10.3847/1538-4357/ac66dd",
    journal = "Astrophys. J.",
    volume = "931",
    number = "1",
    pages = "25",
    year = "2022"
}

@article{Tsunetoe:2024uzh,
    author = "Tsunetoe, Yuh and Kawashima, Tomohisa and Ohsuga, Ken and Mineshige, Shin",
    title = "{Survey of non-thermal electrons around supermassive black holes through polarization flips}",
    eprint = "2409.00171",
    archivePrefix = "arXiv",
    primaryClass = "astro-ph.HE",
    doi = "10.1093/pasj/psae083",
    journal = "Publ. Astron. Soc. Jap.",
    volume = "76",
    number = "6",
    pages = "1211--1227",
    year = "2024"
}

@article{Blandford:1977ds,
    author = "Blandford, R. D. and Znajek, R. L.",
    title = "{Electromagnetic extractions of energy from Kerr black holes}",
    doi = "10.1093/mnras/179.3.433",
    journal = "Mon. Not. Roy. Astron. Soc.",
    volume = "179",
    pages = "433--456",
    year = "1977"
}

@article{Park:2022vzb,
    author = "Park, Jongho and Algaba, Juan Carlos",
    title = "{Polarization Observations of AGN Jets: Past and Future}",
    eprint = "2210.13819",
    archivePrefix = "arXiv",
    primaryClass = "astro-ph.HE",
    doi = "10.3390/galaxies10050102",
    journal = "Galaxies",
    volume = "10",
    number = "5",
    pages = "102",
    year = "2022"
}

@article{Kovalev:2007hy,
    author = "Kovalev, Y. Y. and Lister, M. L. and Homan, D. C. and Kellermann, K. I.",
    title = "{The Inner Jet of the Radio Galaxy M87}",
    eprint = "0708.2695",
    archivePrefix = "arXiv",
    primaryClass = "astro-ph",
    doi = "10.1086/522603",
    journal = "Astrophys. J. Lett.",
    volume = "668",
    pages = "L27",
    year = "2007"
}

@article{napier1994very,
  title={The very long baseline array},
  author={Napier, Peter J and Bagri, Durgadas S and Clark, Barry G and Rogers, Alan EE and Romney, Jonathan D and Thompson, A Richard and Walker, R Craig},
  journal={Proceedings of the IEEE},
  volume={82},
  number={5},
  pages={658--672},
  year={1994},
  publisher={IEEE}
}

@article{hada2011origin,
  title={An origin of the radio jet in M87 at the location of the central black hole},
  author={Hada, Kazuhiro and Doi, Akihiro and Kino, Motoki and Nagai, Hiroshi and Hagiwara, Yoshiaki and Kawaguchi, Noriyuki},
  journal={Nature},
  volume={477},
  number={7363},
  pages={185--187},
  year={2011},
  publisher={Nature Publishing Group UK London}
}

@article{Doeleman:2012zc,
    author = "Doeleman, Sheperd S. and others",
    title = "{Jet Launching Structure Resolved Near the Supermassive Black Hole in M87}",
    eprint = "1210.6132",
    archivePrefix = "arXiv",
    primaryClass = "astro-ph.HE",
    doi = "10.1126/science.1224768",
    journal = "Science",
    volume = "338",
    pages = "355",
    year = "2012"
}

@article{Kim:2018hul,
    author = "Kim, J. -Y. and Krichbaum, T. P. and Lu, R. -S. and Ros, E. and Bach, U. and Bremer, M. and de Vicente, P. and Lindqvist, M. and Zensus, J. A.",
    title = "{The limb-brightened jet of M87 down to 7 Schwarzschild radii scale}",
    eprint = "1805.02478",
    archivePrefix = "arXiv",
    primaryClass = "astro-ph.GA",
    doi = "10.1051/0004-6361/201832921",
    journal = "Astron. Astrophys.",
    volume = "616",
    pages = "A188",
    year = "2018"
}

@article{CraigWalker:2018vam,
    author = "Craig Walker, R. and Hardee, Phillip E. and Davies, Frederick B. and Ly, Chun and Junor, William",
    title = "{The Structure and Dynamics of the Subparsec Jet in M87 Based on 50 VLBA Observations over 17 Years at 43 GHz}",
    eprint = "1802.06166",
    archivePrefix = "arXiv",
    primaryClass = "astro-ph.HE",
    doi = "10.3847/1538-4357/aaafcc",
    journal = "Astrophys. J.",
    volume = "855",
    number = "2",
    pages = "128",
    year = "2018"
}

@article{EventHorizonTelescope:2019dse,
    author = "Akiyama, Kazunori and others",
    collaboration = "Event Horizon Telescope",
    title = "{First M87 Event Horizon Telescope Results. I. The Shadow of the Supermassive Black Hole}",
    eprint = "1906.11238",
    archivePrefix = "arXiv",
    primaryClass = "astro-ph.GA",
    doi = "10.3847/2041-8213/ab0ec7",
    journal = "Astrophys. J. Lett.",
    volume = "875",
    pages = "L1",
    year = "2019"
}

@article{macdonald1968observations,
  title={Observations of the structure of radio sources in the 3C catalogue—I},
  author={Macdonald, GH and Kenderdine, S and Neville, Ann C},
  journal={Monthly Notices of the Royal Astronomical Society},
  volume={138},
  number={3},
  pages={259--311},
  year={1968},
  publisher={Oxford University Press Oxford, UK}
}

@article{Attridge:1999fw,
    author = "Attridge, Joanne M. and Roberts, David H. and Wardle, John F. C.",
    title = "{Radio jet-ambient medium interactions on parsec scales in the blazar 1055+018}",
    eprint = "astro-ph/9903330",
    archivePrefix = "arXiv",
    reportNumber = "BRX-TH-455",
    doi = "10.1086/312078",
    journal = "Astrophys. J. Lett.",
    volume = "518",
    pages = "L87--L90",
    year = "1999"
}

@article{biretta1999hubble,
  title={Hubble Space Telescope observations of superluminal motion in the M87 jet},
  author={Biretta, JA and Sparks, WB and Macchetto, F},
  journal={The Astrophysical Journal},
  volume={520},
  number={2},
  pages={621},
  year={1999},
  publisher={IOP Publishing}
}

@article{junor1999formation,
  title={Formation of the radio jet in M87 at 100 Schwarzschild radii from the central black hole},
  author={Junor, William and Biretta, John A and Livio, Mario},
  journal={Nature},
  volume={401},
  number={6756},
  pages={891--892},
  year={1999},
  publisher={Nature Publishing Group UK London}
}

@article{Lyutikov:2004kn,
    author = "Lyutikov, Maxim and Pariev, Vladimir I. and Gabuzda, Denise C.",
    title = "{Polarization and structure of relativistic parsec-scale AGN jets}",
    eprint = "astro-ph/0406144",
    archivePrefix = "arXiv",
    reportNumber = "SLAC-PUB-10491",
    doi = "10.1111/j.1365-2966.2005.08954.x",
    journal = "Mon. Not. Roy. Astron. Soc.",
    volume = "360",
    pages = "869--891",
    year = "2005"
}

@article{pushkarev2005spine,
  title={Spine--sheath polarization structures in four active galactic nuclei jets},
  author={Pushkarev, AB and Gabuzda, DC and Vetukhnovskaya, Yu N and Yakimov, VE},
  journal={Monthly Notices of the Royal Astronomical Society},
  volume={356},
  number={3},
  pages={859--871},
  year={2005},
  publisher={Blackwell Science Ltd Oxford, UK}
}

@article{Hovatta:2008mj,
    author = "Hovatta, T. and Valtaoja, E. and Tornikoski, M. and Lahteenmaki, A.",
    title = "{Doppler factors, Lorentz factors and viewing angles for quasars, BL Lacertae objects and radio galaxies}",
    eprint = "0811.4278",
    archivePrefix = "arXiv",
    primaryClass = "astro-ph",
    doi = "10.1051/0004-6361:200811150",
    journal = "Astron. Astrophys.",
    volume = "494",
    pages = "527",
    year = "2009"
}

@article{o2009three,
  title={Three-dimensional magnetic field structure of six parsec-scale active galactic nuclei jets},
  author={O'Sullivan, Shane P and Gabuzda, Denise C},
  journal={Monthly Notices of the Royal Astronomical Society},
  volume={393},
  number={2},
  pages={429--456},
  year={2009},
  publisher={Blackwell Publishing Ltd Oxford, UK}
}

@article{Asada:2011dr,
    author = "Asada, Keiichi and Nakamura, Masanori",
    title = "{The Structure of the M87 Jet: A Transition from Parabolic to Conical Streamlines}",
    eprint = "1110.1793",
    archivePrefix = "arXiv",
    primaryClass = "astro-ph.HE",
    doi = "10.1088/2041-8205/745/2/L28",
    journal = "Astrophys. J. Lett.",
    volume = "745",
    pages = "L28",
    year = "2012"
}

@article{Hada:2013yla,
    author = "Hada, Kazuhiro and Kino, Motoki and Doi, Akihiro and Nagai, Hiroshi and Honma, Mareki and Hagiwara, Yoshiaki and Giroletti, Marcello and Giovannini, Gabriele and Kawaguchi, Noriyuki",
    title = "{Innermost collimation structure of the M87 jet down to {\textasciitilde}ten Schwarzschild radii}",
    eprint = "1308.1411",
    archivePrefix = "arXiv",
    primaryClass = "astro-ph.CO",
    doi = "10.1088/0004-637X/775/1/70",
    journal = "Astrophys. J.",
    volume = "775",
    pages = "70",
    year = "2013"
}

@article{Gabuzda:2014tza,
    author = "Gabuzda, Denise C. and Reichstein, Andrea and O'Neill, Eamonn L.",
    title = "{Are spine{\textendash}sheath polarization structures in the jets of active galactic nuclei associated with helical magnetic fields?}",
    eprint = "1410.6653",
    archivePrefix = "arXiv",
    primaryClass = "astro-ph.GA",
    doi = "10.1093/mnras/stu1381",
    journal = "Mon. Not. Roy. Astron. Soc.",
    volume = "444",
    number = "1",
    pages = "172--184",
    year = "2014"
}

@article{Hada:2015okc,
    author = "Hada, Kazuhiro and others",
    title = "{High-Sensitivity 86GHz (3.5mm) VLBI Observations of M87: Deep Imaging of the Jet Base at a 10 Schwarzschild-Radius Resolution}",
    eprint = "1512.03783",
    archivePrefix = "arXiv",
    primaryClass = "astro-ph.HE",
    doi = "10.3847/0004-637X/817/2/131",
    journal = "Astrophys. J.",
    volume = "817",
    number = "2",
    pages = "131",
    year = "2016"
}

@article{Homan:2014uea,
    author = "Homan, D. C. and Lister, M. L. and Kovalev, Y. Y. and Pushkarev, A. B. and Savolainen, T. and Kellermann, K. I. and Richards, J. L. and Ros, E.",
    title = "{MOJAVE XII: Acceleration and Collimation of Blazar Jets on Parsec Scales}",
    eprint = "1410.8502",
    archivePrefix = "arXiv",
    primaryClass = "astro-ph.HE",
    doi = "10.1088/0004-637X/798/2/134",
    journal = "Astrophys. J.",
    volume = "798",
    number = "2",
    pages = "134",
    year = "2015"
}

@article{lee2016interferometric,
  title={Interferometric monitoring of gamma-ray bright AGNs. I. the results of single-epoch multifrequency observations},
  author={Lee, Sang-Sung and Wajima, Kiyoaki and Algaba, Juan-Carlos and Zhao, Guang-Yao and Hodgson, Jeffrey A and Kim, Dae-Won and Park, Jongho and Kim, Jae-Young and Miyazaki, Atsushi and Byun, Do-Young and others},
  journal={The Astrophysical Journal Supplement Series},
  volume={227},
  number={1},
  pages={8},
  year={2016},
  publisher={IOP Publishing}
}

@article{Mertens:2016rhi,
    author = "Mertens, F. and Lobanov, A. P. and Walker, R. C. and Hardee, P. E.",
    title = "{Kinematics of the jet in M 87 on scales of 100{\textendash}1000 Schwarzschild radii}",
    eprint = "1608.05063",
    archivePrefix = "arXiv",
    primaryClass = "astro-ph.HE",
    doi = "10.1051/0004-6361/201628829",
    journal = "Astron. Astrophys.",
    volume = "595",
    pages = "A54",
    year = "2016"
}

@article{Pushkarev:2017fbk,
    author = "Pushkarev, A. B. and Kovalev, Y. Y. and Lister, M. L. and Savolainen, T.",
    title = "{MOJAVE {\textendash} XIV. Shapes and opening angles of AGN jets}",
    eprint = "1705.02888",
    archivePrefix = "arXiv",
    primaryClass = "astro-ph.HE",
    doi = "10.1093/mnras/stx854",
    journal = "Mon. Not. Roy. Astron. Soc.",
    volume = "468",
    number = "4",
    pages = "4992--5003",
    year = "2017"
}

@article{Kutkin:2018qmn,
    author = "Kutkin, Alexander and Pashchenko, Ilya and Sokolovsky, Kirill and Kovalev, Yuri Y. and Aller, Margo and Aller, Hugh",
    title = "{Opacity, variability and kinematics of AGN jets}",
    eprint = "1809.05536",
    archivePrefix = "arXiv",
    primaryClass = "astro-ph.GA",
    doi = "10.1093/mnras/stz885",
    journal = "Mon. Not. Roy. Astron. Soc.",
    volume = "486",
    number = "1",
    pages = "430--439",
    year = "2019"
}

@article{Park:2020mbb,
    author = "Park, Jongho and Hada, Kazuhiro and Nakamura, Masanori and Asada, Keiichi and Zhao, Guang-Yao and Kino, Motoki",
    title = "{Jet collimation and acceleration in the giant radio galaxy NGC 315}",
    eprint = "2012.14154",
    archivePrefix = "arXiv",
    primaryClass = "astro-ph.HE",
    doi = "10.3847/1538-4357/abd6ee",
    journal = "Astrophys. J.",
    volume = "909",
    number = "1",
    pages = "76",
    year = "2021"
}

@article{EventHorizonTelescope:2020dlu,
    author = "Kim, Jae-Young and others",
    collaboration = "Event Horizon Telescope",
    title = "{Event Horizon Telescope imaging of the archetypal blazar 3C 279 at an extreme 20 microarcsecond resolution}",
    reportNumber = "FERMILAB-PUB-20-537-V",
    doi = "10.1051/0004-6361/202037493",
    journal = "Astron. Astrophys.",
    volume = "640",
    pages = "A69",
    year = "2020"
}

@article{Burd:2021sjw,
    author = "Burd, P. R. and Kadler, M. and Mannheim, K. and Baczko, A. -K. and Ringholz, J. and Ros, E.",
    title = "{Dual-high-frequency VLBI study of blazar-jet brightness-temperature gradients and collimation profiles}",
    eprint = "2112.04403",
    archivePrefix = "arXiv",
    primaryClass = "astro-ph.HE",
    doi = "10.1051/0004-6361/202142363",
    journal = "Astron. Astrophys.",
    volume = "660",
    pages = "A1",
    year = "2022"
}

@article{Lu:2023bbn,
    author = "Lu, Ru-Sen and others",
    title = "{A ring-like accretion structure in M87 connecting its black hole and jet}",
    eprint = "2304.13252",
    archivePrefix = "arXiv",
    primaryClass = "astro-ph.HE",
    doi = "10.1038/s41586-023-05843-w",
    journal = "Nature",
    volume = "616",
    number = "7958",
    pages = "686--690",
    year = "2023"
}

@article{Baghel:2023equ,
    author = "Baghel, Janhavi and Kharb, P. and Hovatta, T. and Gulati, S. and Lindfors, E. and S., Silpa",
    title = "{A Kpc-scale radio polarization study of PG BL Lacs with the uGMRT}",
    eprint = "2310.08989",
    archivePrefix = "arXiv",
    primaryClass = "astro-ph.GA",
    doi = "10.1093/mnras/stad3173",
    journal = "Mon. Not. Roy. Astron. Soc.",
    volume = "527",
    number = "1",
    pages = "672--688",
    year = "2023"
}

@article{EventHorizonTelescope:2025uqi,
    author = "Saurabh and others",
    collaboration = "Event Horizon Telescope",
    title = "{Probing jet base emission of M87* with the 2021 Event Horizon Telescope observations}",
    eprint = "2512.08970",
    archivePrefix = "arXiv",
    primaryClass = "astro-ph.HE",
    month = "12",
    year = "2025"
}

@article{Kovalev:2025kxf,
    author = "Kovalev, Y. Y. and Pushkarev, A. B. and Gomez, J. L. and Homan, D. C. and Lister, M. L. and Livingston, J. D. and Pashchenko, I. N. and Plavin, A. V. and Savolainen, T. and Troitsky, S. V.",
    title = "{Looking into the jet cone of the neutrino-associated very high-energy blazar PKS 1424+240}",
    eprint = "2504.09287",
    archivePrefix = "arXiv",
    primaryClass = "astro-ph.HE",
    reportNumber = "INR-TH-2025-004",
    doi = "10.1051/0004-6361/202555400",
    journal = "Astron. Astrophys.",
    volume = "700",
    pages = "L12",
    year = "2025"
}

@article{lyutikov2003polarization,
  title={Polarization of prompt gamma-ray burst emission: evidence for electromagnetically dominated outflow},
  author={Lyutikov, Maxim and Pariev, VI and Blandford, Roger D},
  journal={The Astrophysical Journal},
  volume={597},
  number={2},
  pages={998--1009},
  year={2003}
}

@article{Broderick:2008qf,
    author = "Broderick, Avery and Loeb, Abraham",
    title = "{Imaging the Black Hole Silhouette of M87: Implications for Jet Formation and Black Hole Spin}",
    eprint = "0812.0366",
    archivePrefix = "arXiv",
    primaryClass = "astro-ph",
    doi = "10.1088/0004-637X/697/2/1164",
    journal = "Astrophys. J.",
    volume = "697",
    pages = "1164--1179",
    year = "2009"
}

@article{Kawashima:2020rmr,
    author = "Kawashima, Tomohisa and Toma, Kenji and Kino, Motoki and Akiyama, Kazunori and Nakamura, Masanori and Moriyama, Kotaro",
    title = "{A Jet-bases Emission Model of the EHT2017 Image of M87*}",
    eprint = "2009.08641",
    archivePrefix = "arXiv",
    primaryClass = "astro-ph.HE",
    doi = "10.3847/1538-4357/abd5bb",
    journal = "Astrophys. J.",
    volume = "909",
    number = "2",
    pages = "168",
    year = "2021"
}

@article{Chael:2023pwp,
    author = "Chael, Andrew and Lupsasca, Alexandru and Wong, George N. and Quataert, Eliot",
    title = "{Black Hole Polarimetry I. A Signature of Electromagnetic Energy Extraction}",
    eprint = "2307.06372",
    archivePrefix = "arXiv",
    primaryClass = "astro-ph.HE",
    doi = "10.3847/1538-4357/acf92d",
    journal = "Astrophys. J.",
    volume = "958",
    number = "1",
    pages = "65",
    year = "2023"
}

@article{Gelles:2024tpz,
    author = "Gelles, Zachary and Chael, Andrew and Quataert, Eliot",
    title = "{Signatures of Black Hole Spin and Plasma Acceleration in Jet Polarimetry}",
    eprint = "2410.00954",
    archivePrefix = "arXiv",
    primaryClass = "astro-ph.HE",
    doi = "10.3847/1538-4357/adb1aa",
    journal = "Astrophys. J.",
    volume = "981",
    number = "2",
    pages = "204",
    year = "2025"
}

@article{Gelles:2026mxg,
    author = "Gelles, Zachary and Chael, Andrew and Quataert, Eliot",
    title = "{Signatures of Black Hole Spin and Plasma Acceleration in Jet Polarimetry II: Off-Axis Jets}",
    eprint = "2601.13307",
    archivePrefix = "arXiv",
    primaryClass = "astro-ph.HE",
    month = "1",
    year = "2026"
}

@article{Tsunetoe:2020pyz,
    author = "Tsunetoe, Yuh and Mineshige, Shin and Ohsuga, Ken and Kawashima, Tomohisa and Akiyama, Kazunori",
    title = "{Polarization imaging of M 87 jets by general relativistic radiative transfer calculation based on GRMHD simulations}",
    eprint = "2002.00954",
    archivePrefix = "arXiv",
    primaryClass = "astro-ph.HE",
    doi = "10.1093/pasj/psaa008",
    journal = "Publ. Astron. Soc. Jap.",
    volume = "72",
    number = "2",
    pages = "32",
    year = "2020"
}

@article{Tsunetoe:2024eew,
    author = "Tsunetoe, Yuh and Narayan, Ramesh and Ricarte, Angelo",
    title = "{Jet Archaeology and Forecasting: Image Variability and Magnetic Field Configuration}",
    eprint = "2411.08116",
    archivePrefix = "arXiv",
    primaryClass = "astro-ph.HE",
    doi = "10.3847/1538-4357/adbdcf",
    journal = "Astrophys. J.",
    volume = "983",
    number = "1",
    pages = "77",
    year = "2025"
}

@article{Tsunetoe:2025crz,
    author = "Tsunetoe, Yuh and Pesce, Dominic W. and Narayan, Ramesh and Chael, Andrew and Gelles, Zachary and Gammie, Charles F. and Quataert, Eliot and Palumbo, Daniel C. M.",
    title = "{Limb-brightened Jet in M87 from Anisotropic Nonthermal Electrons}",
    eprint = "2501.14862",
    archivePrefix = "arXiv",
    primaryClass = "astro-ph.HE",
    doi = "10.3847/1538-4357/adc37a",
    journal = "Astrophys. J.",
    volume = "984",
    number = "1",
    pages = "35",
    year = "2025"
}

@article{Jones:2026fbg,
    author = "Jones, Dashon Michel and Anantua, Richard and Emami, Razieh and Lujan, Nate",
    title = "{Probing Axions with Relativistic Jet Polarimetry}",
    eprint = "2603.03244",
    archivePrefix = "arXiv",
    primaryClass = "astro-ph.HE",
    month = "3",
    year = "2026"
}

@article{bekenstein1978new,
  title={New conservation laws in general-relativistic magnetohydrodynamics},
  author={Bekenstein, Jacob D and Oron, Eliezer},
  journal={Physical Review D},
  volume={18},
  number={6},
  pages={1809},
  year={1978},
  publisher={APS}
}

@article{thorne1982electrodynamics,
  title={Electrodynamics in curved spacetime: 3+ 1 formulation},
  author={Thorne, Kip S and Macdonald, Douglas},
  journal={Monthly Notices of the Royal Astronomical Society},
  volume={198},
  number={2},
  pages={339--343},
  year={1982},
  publisher={Oxford University Press}
}

@article{fendt1995collimation,
  title={On the collimation of stellar magnetospheres to jets. I. Relativistic force-free 2D equilibrium.},
  author={Camenzind, M and Appl, S},
  journal={Astronomy and Astrophysics, v. 300, p. 791},
  volume={300},
  pages={791},
  year={1995}
}

@article{fendt1996collimated,
  title={On collimated stellar jet magnetospheres. II. Dynamical structure of collimating wind flows.},
  author={Fendt, Christian and Camenzind, M},
  journal={Astronomy and Astrophysics, v. 313, p. 591-604},
  volume={313},
  pages={591--604},
  year={1996}
}

@article{Gammie:2003rj,
    author = "Gammie, Charles F. and McKinney, Jonathan C. and Toth, Gabor",
    title = "{HARM: A Numerical scheme for general relativistic magnetohydrodynamics}",
    eprint = "astro-ph/0301509",
    archivePrefix = "arXiv",
    doi = "10.1086/374594",
    journal = "Astrophys. J.",
    volume = "589",
    pages = "444--457",
    year = "2003"
}

@article{McKinney:2006tf,
    author = "McKinney, Jonathan C.",
    title = "{General relativistic magnetohydrodynamic simulations of jet formation and large-scale propagation from black hole accretion systems}",
    eprint = "astro-ph/0603045",
    archivePrefix = "arXiv",
    doi = "10.1111/j.1365-2966.2006.10256.x",
    journal = "Mon. Not. Roy. Astron. Soc.",
    volume = "368",
    pages = "1561--1582",
    year = "2006"
}

@article{Davelaar:2023dhl,
    author = "Davelaar, Jordy and Ripperda, Bart and Sironi, Lorenzo and Philippov, Alexander A. and Olivares, Hector and Porth, Oliver and Berg, Bram van den and Bronzwaer, Thomas and Chatterjee, Koushik and Liska, Matthew",
    title = "{Synchrotron Polarization Signatures of Surface Waves in Supermassive Black Hole Jets}",
    eprint = "2309.07963",
    archivePrefix = "arXiv",
    primaryClass = "astro-ph.HE",
    doi = "10.3847/2041-8213/ad0b79",
    journal = "Astrophys. J. Lett.",
    volume = "959",
    number = "1",
    pages = "L3",
    year = "2023"
}

@article{Song:2025mhj,
    author = "Song, Yu and Hou, Yehui and Huang, Lei and Chen, Bin",
    title = "{Hybrid Black Hole-Disk Driven Jets: Steady Axisymmetric Ideal MHD Modeling}",
    eprint = "2507.23281",
    archivePrefix = "arXiv",
    primaryClass = "astro-ph.HE",
    month = "7",
    year = "2025"
}

@article{Tomimatsu:2003uz,
    author = "Tomimatsu, Akira and Takahashi, Masaaki",
    title = "{Relativistic acceleration of magnetically driven jets}",
    eprint = "astro-ph/0303509",
    archivePrefix = "arXiv",
    doi = "10.1086/375579",
    journal = "Astrophys. J.",
    volume = "592",
    pages = "321--331",
    year = "2003"
}

@article{Pu:2017akw,
    author = "Pu, Hung-Yi and Wu, Kinwah and Younsi, Ziri and Asada, Keiichi and Mizuno, Yosuke and Nakamura, Masanori",
    title = "{Observable Emission Features of Black Hole GRMHD Jets on Event Horizon Scales}",
    eprint = "1707.07023",
    archivePrefix = "arXiv",
    primaryClass = "astro-ph.HE",
    doi = "10.3847/1538-4357/aa8136",
    journal = "Astrophys. J.",
    volume = "845",
    number = "2",
    pages = "160",
    year = "2017"
}

@article{znajek1977black,
  title={Black hole electrodynamics and the Carter tetrad},
  author={Znajek, Roman L},
  journal={Monthly Notices of the Royal Astronomical Society},
  volume={179},
  number={3},
  pages={457--472},
  year={1977},
  publisher={Oxford University Press Oxford, UK}
}

@article{Kino:2022xme,
    author = "Kino, Motoki and Takahashi, Masaaki and Kawashima, Tomohisa and Park, Jongho and Hada, Kazuhiro and Ro, Hyunwook and Cui, Yuzhu",
    title = "{Implications from the Velocity Profile of the M87 Jet: A Possibility of a Slowly Rotating Black Hole Magnetosphere}",
    eprint = "2209.07264",
    archivePrefix = "arXiv",
    primaryClass = "astro-ph.HE",
    doi = "10.3847/1538-4357/ac8c2f",
    journal = "Astrophys. J.",
    volume = "939",
    number = "2",
    pages = "83",
    year = "2022"
}

@article{Cruz-Osorio:2021cob,
    author = "Cruz-Osorio, Alejandro and Fromm, Christian M. and Mizuno, Yosuke and Nathanail, Antonios and Younsi, Ziri and Porth, Oliver and Davelaar, Jordy and Falcke, Heino and Kramer, Michael and Rezzolla, Luciano",
    title = "{State-of-the-art energetic and morphological modelling of the launching site of the M87 jet}",
    eprint = "2111.02517",
    archivePrefix = "arXiv",
    primaryClass = "astro-ph.HE",
    doi = "10.1038/s41550-021-01506-w",
    journal = "Nature Astron.",
    volume = "6",
    number = "1",
    pages = "103--108",
    year = "2022"
}

@article{Yang:2024kpz,
    author = "Yang, Hai and Yuan, Feng and Li, Hui and Mizuno, Yosuke and Guo, Fan and Lu, Rusen and Ho, Luis C. and Lin, Xi and Zdziarski, Andrzej A. and Wang, Jieshuang",
    title = "{Modeling the inner part of the jet in M87: Confronting jet morphology with theory}",
    eprint = "2403.15950",
    archivePrefix = "arXiv",
    primaryClass = "astro-ph.HE",
    doi = "10.1126/sciadv.adn3544",
    journal = "Sci. Adv.",
    volume = "10",
    number = "12",
    pages = "adn3544",
    year = "2024"
}

@article{Moscibrodzka:2015pda,
    author = "Moscibrodzka, Monika and Falcke, Heino and Shiokawa, Hotaka",
    title = "{General relativistic magnetohydrodynamical simulations of the jet in M 87}",
    eprint = "1510.07243",
    archivePrefix = "arXiv",
    primaryClass = "astro-ph.HE",
    doi = "10.1051/0004-6361/201526630",
    journal = "Astron. Astrophys.",
    volume = "586",
    pages = "A38",
    year = "2016"
}

@article{EventHorizonTelescope:2021bee,
    author = "Akiyama, Kazunori and others",
    collaboration = "Event Horizon Telescope",
    title = "{First M87 Event Horizon Telescope Results. VII. Polarization of the Ring}",
    eprint = "2105.01169",
    archivePrefix = "arXiv",
    primaryClass = "astro-ph.HE",
    reportNumber = "FERMILAB-PUB-21-849-PPD",
    doi = "10.3847/2041-8213/abe71d",
    journal = "Astrophys. J. Lett.",
    volume = "910",
    number = "1",
    pages = "L12",
    year = "2021"
}

@article{EventHorizonTelescope:2021srq,
    author = "Akiyama, Kazunori and others",
    collaboration = "Event Horizon Telescope",
    title = "{First M87 Event Horizon Telescope Results. VIII. Magnetic Field Structure near The Event Horizon}",
    eprint = "2105.01173",
    archivePrefix = "arXiv",
    primaryClass = "astro-ph.HE",
    reportNumber = "FERMILAB-PUB-21-850-PPD",
    doi = "10.3847/2041-8213/abe4de",
    journal = "Astrophys. J. Lett.",
    volume = "910",
    number = "1",
    pages = "L13",
    year = "2021"
}

@article{Papoutsis:2022kzp,
    author = {Papoutsis, Evan and Baub\"ock, Michi and Chang, Dominic and Gammie, Charles F.},
    title = "{Jets and Rings in Images of Spinning Black Holes}",
    eprint = "2212.06281",
    archivePrefix = "arXiv",
    primaryClass = "astro-ph.HE",
    doi = "10.3847/1538-4357/acafe3",
    journal = "Astrophys. J.",
    volume = "944",
    number = "1",
    pages = "55",
    year = "2023"
}

@article{Blandford:2018iot,
    author = "Blandford, Roger and Meier, David and Readhead, Anthony",
    title = "{Relativistic Jets from Active Galactic Nuclei}",
    eprint = "1812.06025",
    archivePrefix = "arXiv",
    primaryClass = "astro-ph.HE",
    doi = "10.1146/annurev-astro-081817-051948",
    journal = "Ann. Rev. Astron. Astrophys.",
    volume = "57",
    pages = "467--509",
    year = "2019"
}

@article{Tchekhovskoy:2011zx,
    author = "Tchekhovskoy, Alexander and Narayan, Ramesh and McKinney, Jonathan C.",
    title = "{Efficient Generation of Jets from Magnetically Arrested Accretion on a Rapidly Spinning Black Hole}",
    eprint = "1108.0412",
    archivePrefix = "arXiv",
    primaryClass = "astro-ph.HE",
    doi = "10.1111/j.1745-3933.2011.01147.x",
    journal = "Mon. Not. Roy. Astron. Soc.",
    volume = "418",
    pages = "L79--L83",
    year = "2011"
}

@article{McKinney:2012vh,
    author = "McKinney, Jonathan C. and Tchekhovskoy, Alexander and Blandford, Roger D.",
    title = "{General Relativistic Magnetohydrodynamic Simulations of Magnetically Choked Accretion Flows around Black Holes}",
    eprint = "1201.4163",
    archivePrefix = "arXiv",
    primaryClass = "astro-ph.HE",
    reportNumber = "SLAC-PUB-14950",
    doi = "10.1111/j.1365-2966.2012.21074.x",
    journal = "Mon. Not. Roy. Astron. Soc.",
    volume = "423",
    pages = "3083",
    year = "2012"
}

@article{Gralla:2014yja,
    author = "Gralla, Samuel E. and Jacobson, Ted",
    title = "{Spacetime approach to force-free magnetospheres}",
    eprint = "1401.6159",
    archivePrefix = "arXiv",
    primaryClass = "astro-ph.HE",
    doi = "10.1093/mnras/stu1690",
    journal = "Mon. Not. Roy. Astron. Soc.",
    volume = "445",
    number = "3",
    pages = "2500--2534",
    year = "2014"
}

@article{Huang:2019wqv,
    author = "Huang, Lei and Pan, Zhen and Yu, Cong",
    title = "{Towards a Full MHD Jet Model of Spinning Black Holes--I: Framework and a split monopole example}",
    eprint = "1906.05454",
    archivePrefix = "arXiv",
    primaryClass = "astro-ph.HE",
    doi = "10.3847/1538-4357/ab2909",
    month = "6",
    year = "2019"
}

@article{komissarov2004electrodynamics,
  title={Electrodynamics of black hole magnetospheres},
  author={Komissarov, SS},
  journal={Monthly Notices of the Royal Astronomical Society},
  volume={350},
  number={2},
  pages={427--448},
  year={2004},
  publisher={Blackwell Science Ltd Oxford, UK}
}

@article{goldreich1970stellar,
  title={Stellar winds},
  author={Goldreich, Peter and Julian, William H},
  journal={Astrophysical Journal, vol. 160, p. 971},
  volume={160},
  pages={971},
  year={1970}
}

@article{michel1969relativistic,
  title={Relativistic stellar-wind torques},
  author={Michel, FC},
  journal={Astrophysical Journal, vol. 158, p. 727},
  volume={158},
  pages={727},
  year={1969}
}

@article{Nitta:1991ui,
    author = "Nitta, Shin-ya and Takahashi, Masaaki and Tomimatsu, Akira",
    title = "{Effects of magnetohydrodynamic accretion flows on global structure of a Kerr black hole magnetosphere}",
    reportNumber = "DPNU-91-13",
    doi = "10.1103/PhysRevD.44.2295",
    journal = "Phys. Rev. D",
    volume = "44",
    pages = "2295--2305",
    year = "1991"
}

@article{sikora2000pair,
  title={On pair content and variability of subparsec jets in quasars},
  author={Sikora, Marek and Madejski, Greg},
  journal={The Astrophysical Journal},
  volume={534},
  number={1},
  pages={109--113},
  year={2000}
}

@article{lightman1987pair,
  title={Pair production and Compton scattering in compact sources and comparison to observations of active galactic nuclei},
  author={Lightman, Alan P and Zdziarski, Andrzej A},
  journal={Astrophysical Journal, Part 1 (ISSN 0004-637X), vol. 319, Aug. 15, 1987, p. 643-661.},
  volume={319},
  pages={643--661},
  year={1987}
}

@article{Blandford:1995yf,
    author = "Blandford, R. D. and Levinson, A.",
    title = "{Pair cascades in extragalactic jets. 1: Gamma rays}",
    doi = "10.1086/175338",
    journal = "Astrophys. J.",
    volume = "441",
    pages = "79--95",
    year = "1995"
}

@article{Broderick:2015swa,
    author = "Broderick, Avery E and Tchekhovskoy, Alexander",
    title = "{Horizon-Scale Lepton Acceleration in Jets: Explaining the Compact Radio Emission in M87}",
    eprint = "1506.04754",
    archivePrefix = "arXiv",
    primaryClass = "astro-ph.HE",
    doi = "10.1088/0004-637X/809/1/97",
    journal = "Astrophys. J.",
    volume = "809",
    number = "1",
    pages = "97",
    year = "2015"
}

@article{Levinson:2010fc,
    author = "Levinson, Amir and Rieger, Frank",
    title = "{Variable TeV emission as a manifestation of jet formation in M87?}",
    eprint = "1011.5319",
    archivePrefix = "arXiv",
    primaryClass = "astro-ph.HE",
    doi = "10.1088/0004-637X/730/2/123",
    journal = "Astrophys. J.",
    volume = "730",
    pages = "123",
    year = "2011"
}

@article{sokoloff1998depolarization,
  title={Depolarization and Faraday effects in galaxies},
  author={Sokoloff, DD and Bykov, AA and Shukurov, A and Berkhuijsen, EM and Beck, R and Poezd, AD},
  journal={Monthly Notices of the Royal Astronomical Society},
  volume={299},
  number={1},
  pages={189--206},
  year={1998},
  publisher={The Royal Astronomical Society}
}

@article{Blandford:1982xxl,
    author = "Blandford, R. D. and Payne, D. G.",
    title = "{Hydromagnetic flows from accretion discs and the production of radio jets}",
    doi = "10.1093/mnras/199.4.883",
    journal = "Mon. Not. Roy. Astron. Soc.",
    volume = "199",
    number = "4",
    pages = "883--903",
    year = "1982"
}

@article{takahashi1990magnetohydrodynamic,
  title={Magnetohydrodynamic flows in Kerr geometry-Energy extraction from black holes},
  author={Takahashi, Masaaki and Nitta, Shinya and Tatematsu, Yoshinori and Tomimatsu, Akira},
  journal={Astrophysical Journal, Part 1 (ISSN 0004-637X), vol. 363, Nov. 1, 1990, p. 206-217.},
  volume={363},
  pages={206--217},
  year={1990}
}

@article{McKinney:2006dx,
    author = "McKinney, Jonathan C. and Narayan, Ramesh",
    title = "{Disk-Jet Coupling in Black Hole Accretion Systems I: General Relativistic Magnetohydrodynamical Models}",
    eprint = "astro-ph/0607575",
    archivePrefix = "arXiv",
    doi = "10.1111/j.1365-2966.2006.11301.x",
    journal = "Mon. Not. Roy. Astron. Soc.",
    volume = "375",
    pages = "513--530",
    year = "2007"
}

@article{Tchekhovskoy:2008gq,
    author = "Tchekhovskoy, Alexander and McKinney, Jonathan C. and Narayan, Ramesh",
    title = "{Simulations of Ultrarelativistic Magnetodynamic Jets from Gamma-ray Burst Engines}",
    eprint = "0803.3807",
    archivePrefix = "arXiv",
    primaryClass = "astro-ph",
    doi = "10.1111/j.1365-2966.2008.13425.x",
    journal = "Mon. Not. Roy. Astron. Soc.",
    volume = "388",
    pages = "551",
    year = "2008"
}

@article{Chantry:2022ejm,
    author = "Chantry, L. and Cayatte, V. and Sauty, C. and Vlahakis, N. and Tsinganos, K.",
    title = "{Double flows anchored in a Kerr black hole horizon {\textendash} I. Meridionally self-similar MHD models with loading terms}",
    eprint = "2207.06094",
    archivePrefix = "arXiv",
    primaryClass = "astro-ph.HE",
    doi = "10.1093/mnras/stac1990",
    journal = "Mon. Not. Roy. Astron. Soc.",
    volume = "515",
    number = "3",
    pages = "3796--3817",
    year = "2022"
}

@article{Huang:2020lvl,
    author = "Huang, Lei and Pan, Zhen and Yu, Cong",
    title = "{Toward a Full MHD Jet Model of Spinning Black Holes--II: Kinematics and Application to the M87 Jet}",
    eprint = "2004.01827",
    archivePrefix = "arXiv",
    primaryClass = "astro-ph.HE",
    doi = "10.3847/1538-4357/ab86a3",
    journal = "Astrophys. J.",
    volume = "894",
    number = "1",
    pages = "45",
    year = "2020"
}

@article{Broderick:2003fc,
	author = "Broderick, Avery and Blandford, Roger",
	title = "{Covariant magnetoionic theory. 2. Radiative transfer}",
	eprint = "astro-ph/0311360",
	archivePrefix = "arXiv",
	reportNumber = "SLAC-PUB-10262",
	doi = "10.1111/j.1365-2966.2004.07582.x",
	journal = "Mon. Not. Roy. Astron. Soc.",
	volume = "349",
	pages = "994",
	year = "2004"
}

@article{camenzind1986hydromagnetic,
  title={Hydromagnetic flows from rapidly rotating compact objects. I-Cold relativistic flows from rapid rotators},
  author={Camenzind, M},
  journal={Astronomy and Astrophysics (ISSN 0004-6361), vol. 162, no. 1-2, July 1986, p. 32-44. SNSF-supported research.},
  volume={162},
  pages={32--44},
  year={1986}
}

@article{takahashi1998trans,
  title={Trans-fast MHD winds in a pulsar magnetosphere},
  author={Takahashi, Masaaki and Shibata, Shinpei},
  journal={Publications of the Astronomical Society of Japan},
  volume={50},
  number={2},
  pages={271--283},
  year={1998},
  publisher={OUP}
}

@article{ALMA:2025wvr,
    author = "Goddi, Ciriaco and others",
    collaboration = "ALMA",
    title = "{First polarization study of the M87 jet and active galactic nuclei at submillimeter wavelengths with ALMA}",
    eprint = "2505.10181",
    archivePrefix = "arXiv",
    primaryClass = "astro-ph.GA",
    reportNumber = "FERMILAB-PUB-25-0326-PPD",
    month = "5",
    year = "2025"
}

@article{Vazquez:2003zm,
    author = "Vazquez, Samuel E. and Esteban, Ernesto P.",
    title = "{Strong field gravitational lensing by a Kerr black hole}",
    eprint = "gr-qc/0308023",
    archivePrefix = "arXiv",
    doi = "10.1393/ncb/i2004-10121-y",
    journal = "Nuovo Cim. B",
    volume = "119",
    pages = "489--519",
    year = "2004"
}

@article{Ayzenberg:2023hfw,
    author = "Ayzenberg, D. and others",
    title = "{Fundamental physics opportunities with future ground-based mm/sub-mm VLBI arrays}",
    eprint = "2312.02130",
    archivePrefix = "arXiv",
    primaryClass = "astro-ph.HE",
    doi = "10.1007/s41114-025-00057-0",
    journal = "Living Rev. Rel.",
    volume = "28",
    number = "1",
    pages = "4",
    year = "2025",
    note = "[Erratum: Living Rev.Rel. 28, 7 (2025)]"
}

@article{Johnson:2023ynn,
    author = "Johnson, Michael D. and others",
    title = "{Key Science Goals for the Next-Generation Event Horizon Telescope}",
    eprint = "2304.11188",
    archivePrefix = "arXiv",
    primaryClass = "astro-ph.HE",
    doi = "10.3390/galaxies11030061",
    journal = "Galaxies",
    volume = "11",
    number = "3",
    pages = "61",
    year = "2023"
}

@article{Johnson:2024ttr,
    author = "Johnson, Michael D. and others",
    title = "{The Black Hole Explorer: motivation and vision}",
    eprint = "2406.12917",
    archivePrefix = "arXiv",
    primaryClass = "astro-ph.IM",
    doi = "10.1117/12.3019835",
    journal = "Proc. SPIE Int. Soc. Opt. Eng.",
    volume = "13092",
    pages = "130922D",
    year = "2024"
}

@article{Johnson:2019ljv,
    author = "Johnson, Michael D. and others",
    title = "{Universal interferometric signatures of a black hole{\textquoteright}s photon ring}",
    eprint = "1907.04329",
    archivePrefix = "arXiv",
    primaryClass = "astro-ph.IM",
    doi = "10.1126/sciadv.aaz1310",
    journal = "Sci. Adv.",
    volume = "6",
    number = "12",
    pages = "eaaz1310",
    year = "2020"
}

@article{Dihingia:2021ncv,
    author = "Dihingia, Indu K. and Vaidya, Bhargav and Fendt, Christian",
    title = "{Jets, disc-winds, and oscillations in general relativistic, magnetically driven flows around black hole}",
    eprint = "2105.11468",
    archivePrefix = "arXiv",
    primaryClass = "astro-ph.HE",
    doi = "10.1093/mnras/stab1512",
    journal = "Mon. Not. Roy. Astron. Soc.",
    volume = "505",
    number = "3",
    pages = "3596--3615",
    year = "2021"
}

@article{Emami:2021ick,
    author = "Emami, Razieh and Anantua, Richard and Chael, Andrew A. and Loeb, Abraham",
    title = "{Positron Effects on Polarized Images and Spectra from Jet and Accretion Flow Models of M87* and Sgr A*}",
    eprint = "2101.05327",
    archivePrefix = "arXiv",
    primaryClass = "astro-ph.HE",
    doi = "10.3847/1538-4357/ac2950",
    journal = "Astrophys. J.",
    volume = "923",
    number = "2",
    pages = "272",
    year = "2021"
}

@article{Anantua:2019bna,
    author = "Anantua, Richard and Emami, Razieh and Loeb, Abraham and Chael, Andrew",
    title = "{Determining the Composition of Relativistic Jets from Polarization Maps}",
    eprint = "1909.09230",
    archivePrefix = "arXiv",
    primaryClass = "astro-ph.HE",
    doi = "10.3847/1538-4357/ab9103",
    journal = "Astrophys. J.",
    volume = "896",
    number = "1",
    pages = "30",
    year = "2020"
}

@article{marscher1985models,
  title={Models for high-frequency radio outbursts in extragalactic sources, with application to the early 1983 millimeter-to-infrared flare of 3C 273},
  author={Marscher, Alan P and Gear, Walter Kieran},
  journal={Astrophysical Journal, Part 1 (ISSN 0004-637X), vol. 298, Nov. 1, 1985, p. 114-127.},
  volume={298},
  pages={114--127},
  year={1985}
}

@article{sironi2015relativistic,
  title={Relativistic jets shine through shocks or magnetic reconnection?},
  author={Sironi, Lorenzo and Petropoulou, Maria and Giannios, Dimitrios},
  journal={Monthly Notices of the Royal Astronomical Society},
  volume={450},
  number={1},
  pages={183--191},
  year={2015},
  publisher={The Royal Astronomical Society}
}

@article{Mizuno:2012xd,
    author = "Mizuno, Yosuke and Lyubarsky, Yuri and Nishikawa, Ken-Ichi and Hardee, Philip E.",
    title = "{Three-Dimensional Relativistic Magnetohydrodynamic Simulations of Current-Driven Instability. III. Rotating Relativistic Jets}",
    eprint = "1207.4949",
    archivePrefix = "arXiv",
    primaryClass = "astro-ph.HE",
    doi = "10.1088/0004-637X/757/1/16",
    journal = "Astrophys. J.",
    volume = "757",
    pages = "16",
    year = "2012",
    note = "[Erratum: Astrophys.J. 765, 160 (2013)]"
}

@article{burn1966depolarization,
  title={On the depolarization of discrete radio sources by Faraday dispersion},
  author={Burn, BJ},
  journal={Monthly Notices of the Royal Astronomical Society},
  volume={133},
  number={1},
  pages={67--83},
  year={1966},
  publisher={Oxford University Press Oxford, UK}
}

@article{Walker:1970un,
    author = "Walker, M. and Penrose, R.",
    title = "{On quadratic first integrals of the geodesic equations for type [22] spacetimes}",
    doi = "10.1007/BF01649445",
    journal = "Commun. Math. Phys.",
    volume = "18",
    pages = "265--274",
    year = "1970"
}

@article{cunningham1973optical,
  title={The optical appearance of a star orbiting an extreme Kerr black hole},
  author={Cunningham, Christopher T and Bardeen, James M},
  journal={Astrophysical Journal, Vol. 183, pp. 237-264 (1973)},
  volume={183},
  pages={237--264},
  year={1973}
}

@article{Ball:2018icx,
    author = {Ball, David and Sironi, Lorenzo and {\"O}zel, Feryal},
    title = "{Electron and Proton Acceleration in Trans-Relativistic Magnetic Reconnection: Dependence on Plasma Beta and Magnetization}",
    eprint = "1803.05556",
    archivePrefix = "arXiv",
    primaryClass = "astro-ph.HE",
    doi = "10.3847/1538-4357/aac820",
    journal = "Astrophys. J.",
    volume = "862",
    number = "1",
    pages = "80",
    year = "2018"
}

@article{Lai:2025yeq,
    author = "Lai, Paul C. W. and Li, Kaye J. and Yap, Y. X. Jane and Wu, Kinwah and Kong, Albert K. H.",
    title = "{Spectropolarimetry of synchrotron radiation from relativistic electrons with anisotropic pitch-angle and various energy distributions}",
    eprint = "2509.04572",
    archivePrefix = "arXiv",
    primaryClass = "astro-ph.HE",
    doi = "10.1093/mnras/staf1295",
    month = "9",
    year = "2025"
}

@article{comisso2022ion,
  title={Ion and electron acceleration in fully kinetic plasma turbulence},
  author={Comisso, Luca and Sironi, Lorenzo},
  journal={The Astrophysical Journal Letters},
  volume={936},
  number={2},
  pages={L27},
  year={2022},
  publisher={IOP Publishing}
}

@article{Comisso:2023ygd,
    author = "Comisso, Luca and Jiang, Brian",
    title = "{Pitch-angle Anisotropy Imprinted by Relativistic Magnetic Reconnection}",
    eprint = "2310.17560",
    archivePrefix = "arXiv",
    primaryClass = "astro-ph.HE",
    doi = "10.3847/1538-4357/ad1241",
    journal = "Astrophys. J.",
    volume = "959",
    number = "2",
    pages = "137",
    year = "2023"
}

@article{EventHorizonTelescope:2021btj,
    author = "Akiyama, Kazunori and others",
    collaboration = "Event Horizon Telescope",
    title = "{The Polarized Image of a Synchrotron-emitting Ring of Gas Orbiting a Black Hole}",
    eprint = "2105.01804",
    archivePrefix = "arXiv",
    primaryClass = "astro-ph.HE",
    reportNumber = "FERMILAB-PUB-21-848-PPD",
    doi = "10.3847/1538-4357/abf117",
    journal = "Astrophys. J.",
    volume = "912",
    number = "1",
    pages = "35",
    year = "2021"
}

@article{kulsrud1983mhd,
	title={MHD description of plasma},
	author={Kulsrud, Russell M},
	journal={Handbook of plasma physics},
	volume={1},
	pages={115},
	year={1983},
	publisher={North Holland Publishing Company Amsterdam}
}

@book{1979Lightman,
	title={Lightman Radiative Processes in Astrophysics},
	author={ Rybicki, G. B.  and  Lightman, A. P. },
	publisher={Lightman Radiative Processes in Astrophysics},
	year={1979},
}

@article{Shcherbakov:2010kh,
	author = "Shcherbakov, Roman V. and Huang, Lei",
	title = "{General relativistic polarized radiative transfer: building a dynamics-observations interface}",
	eprint = "1007.4831",
	archivePrefix = "arXiv",
	primaryClass = "astro-ph.HE",
	doi = "10.1111/j.1365-2966.2010.17502.x",
	journal = "Mon. Not. Roy. Astron. Soc.",
	volume = "410",
	pages = "1052",
	year = "2011"
}

@article{Palumbo:2020flt,
    author = "Palumbo, Daniel C. M. and Wong, George N. and Prather, Ben S.",
    title = "{Discriminating Accretion States via Rotational Symmetry in Simulated Polarimetric Images of M87}",
    eprint = "2004.01751",
    archivePrefix = "arXiv",
    primaryClass = "astro-ph.HE",
    doi = "10.3847/1538-4357/ab86ac",
    journal = "Astrophys. J.",
    volume = "894",
    number = "2",
    pages = "156",
    year = "2020"
}

@article{Gralla:2019xty,
    author = "Gralla, Samuel E. and Holz, Daniel E. and Wald, Robert M.",
    title = "{Black Hole Shadows, Photon Rings, and Lensing Rings}",
    eprint = "1906.00873",
    archivePrefix = "arXiv",
    primaryClass = "astro-ph.HE",
    doi = "10.1103/PhysRevD.100.024018",
    journal = "Phys. Rev. D",
    volume = "100",
    number = "2",
    pages = "024018",
    year = "2019"
}

@article{Chen:2025ysv,
    author = "Chen, Chengjia and Pan, Qiyuan and Jing, Jiliang",
    title = "{Near-horizon polarized images of a rotating hairy Horndeski black hole}",
    eprint = "2509.12526",
    archivePrefix = "arXiv",
    primaryClass = "gr-qc",
    month = "9",
    year = "2025"
}

@article{Shen:2026aye,
    author = "Shen, Ye",
    title = "{General Grad-Shafranov Equation}",
    eprint = "2605.08597",
    archivePrefix = "arXiv",
    primaryClass = "gr-qc",
    month = "5",
    year = "2026"
}

@article{Hou:2024qqo,
    author = "Hou, Yehui and Huang, Jiewei and Guo, Minyong and Mizuno, Yosuke and Chen, Bin",
    title = "{Near-horizon Polarization as a Diagnostic of Black Hole Spacetime}",
    eprint = "2409.07248",
    archivePrefix = "arXiv",
    primaryClass = "gr-qc",
    reportNumber = "988 L51",
    doi = "10.3847/2041-8213/adee09",
    journal = "Astrophys. J. Lett.",
    volume = "988",
    number = "2",
    pages = "L51",
    year = "2025"
}

@article{Hou:2023bep,
    author = "Hou, Yehui and Zhang, Zhenyu and Guo, Minyong and Chen, Bin",
    title = "{A new analytical model of magnetofluids surrounding rotating black holes}",
    eprint = "2309.13304",
    archivePrefix = "arXiv",
    primaryClass = "gr-qc",
    doi = "10.1088/1475-7516/2024/02/030",
    journal = "JCAP",
    volume = "02",
    pages = "030",
    year = "2024"
}

@article{Chen:2024jkm,
	author = "Chen, Bin and Hou, Yehui and Song, Yu and Zhang, Zhenyu",
	title = "{Polarization patterns of the hot spots plunging into a Kerr black hole}",
	eprint = "2407.14897",
	archivePrefix = "arXiv",
	primaryClass = "astro-ph.HE",
	doi = "10.1103/PhysRevD.111.083045",
	journal = "Phys. Rev. D",
	volume = "111",
	number = "8",
	pages = "083045",
	year = "2025"
}

@article{Zhang:2024lsf,
    author = "Zhang, Zhenyu and Hou, Yehui and Guo, Minyong and Chen, Bin",
    title = "{Imaging thick accretion disks and jets surrounding black holes}",
    eprint = "2401.14794",
    archivePrefix = "arXiv",
    primaryClass = "astro-ph.HE",
    doi = "10.1088/1475-7516/2024/05/032",
    journal = "JCAP",
    volume = "05",
    pages = "032",
    year = "2024"
}

@article{Zhou:2025moa,
    author = "Zhou, Fan and Huang, Jiewei and Li, Yuehang and Zhang, Zhenyu and Hou, Yehui and Guo, Minyong and Chen, Bin",
    title = "{Nonthermal Synchrotron Emission and Polarization Signatures during Black Hole Flux Eruptions}",
    eprint = "2512.06803",
    archivePrefix = "arXiv",
    primaryClass = "astro-ph.HE",
    doi = "10.3847/1538-4357/ae5e67",
    journal = "Astrophys. J.",
    volume = "1002",
    number = "2",
    pages = "152",
    year = "2026"
}

@article{jetaniso:2026,
    author = "Zhang, Zhenyu and Hou, Yehui and Song, Yu and Yosuke Mizuno and Chen, Bin",
    title = "{In preparation}",
    year = "202x"

 }

@article{Hou:2026snq,
    author = "Hou, Yehui and Huang, Jiewei and Chen, Bin",
    title = "{Distinct Near-Horizon Trend of Synchrotron Polarization in Kerr Spacetime}",
    eprint = "2606.19229",
    archivePrefix = "arXiv",
    primaryClass = "gr-qc",
    month = "6",
    year = "2026"
}
\bibliographystyle{aasjournal}

\end{document}